# On the meaning of feedback parameter, transient climate response, and the greenhouse effect: Basic considerations and the discussion of uncertainties

**Gerhard Kramm**[1] **and Ralph Dlugi**[2]

[1]University of Alaska Fairbanks, Geophysical Institute, 903 Koyukuk Drive, P.O. Box 757320, Fairbanks, AK 99775-7320, USA

[2]Arbeitsgruppe Atmosphärische Prozesse (AGAP), Gernotstraße, D-80804 Munich, Germany

**Abstract**

In this paper we discuss the meaning of feedback parameter, greenhouse effect and transient climate response usually related to the globally averaged energy balance model of Schneider and Mass. After scrutinizing this model and the corresponding planetary radiation balance we state that (a) the this globally averaged energy balance model is flawed by unsuitable physical considerations, (b) the planetary radiation balance for an Earth in the absence of an atmosphere is fraught by the inappropriate assumption of a uniform surface temperature, the so-called radiative equilibrium temperature of about 255 K, and (c) the effect of the radiative anthropogenic forcing, considered as a perturbation to the natural system, is much smaller than the uncertainty involved in the solution of the model of Schneider and Mass. This uncertainty is mainly related to the empirical constants suggested by various authors and used for predicting the emission of infrared radiation by the Earth's skin. Furthermore, after inserting the absorption of solar radiation by atmospheric constituents and the exchange of sensible and latent heat between the Earth and the atmosphere into the model of Schneider and Mass the surface temperatures become appreciably lesser than the radiative equilibrium temperature. Moreover, neither the model of Schneider and Mass nor the Dines-type two-layer energy balance model for the Earth-atmosphere system, both contain the planetary radiation balance for an Earth in the absence of an atmosphere as an asymptotic solution, do not provide evidence for the existence of the so-called atmospheric greenhouse effect if realistic empirical data are used.

**Contents**

---

## 1. Introduction

Recently, Manabe and Stouffer [1] discussed the role of ocean in global warming on the basis of results obtained from coupled ocean-atmosphere models of various complexities. By adopting the globally averaged energy balance model of Schneider and Mass [2] that reads

$$C \frac{\partial T_s}{\partial t} = Q - \lambda\, T_s \quad , \tag{1.1}$$

they also discuss the sensitivity of a global climate system. Here, C is the heat capacity of the system, t is time, $T_s$ is the deviation of the surface temperature from the initial value, Q is a thermal forcing, and $\lambda$ is the so-called feedback parameter [1]. It is stated that the term on the left-hand side describes the change in the heat content of the global climate system expressed by [3]

$$\frac{\partial H}{\partial t} = C \frac{\partial T_s}{\partial t} \tag{1.2}$$

so that Eq. (1.1) becomes

$$\frac{\partial H}{\partial t} = Q - \lambda\, T_s \quad . \tag{1.3}$$

Equations (1.1) and (1.3) are well-know in the literature and may be used to predict the time required for the surface temperature to approach its new equilibrium value in response to a change in climate forcing (e.g., [3, 4]). According to Dickinson [5], this equation may be considered as a very simple global energy balance climate model. It seems, however, that even this simple climate model requires a physical clarification. In addition it is indispensable to point out its limitations.

Recently, the National Research Council (NRC) report provides a definition for the global average surface temperature [6]:

According to the radiative-convective equilibrium concept, the equation for determining global average surface temperature of the planet is

$$\frac{\partial H}{\partial t} = f - \frac{T'}{\lambda *} \qquad (1\text{-}1)$$

where

$$H = \int_{z_b}^{\infty} \rho \, C_p \, T \, dz \qquad (1\text{-}2)$$

is the heat content of the land-ocean-atmosphere system with $\rho$ the density, $C_p$ the specific heat, $T$ the temperature, and $z_b$ the depth to which the heating penetrates. Equation 1-1 describes the change in the heat content, where $f$ is the radiative forcing at the tropopause, $T'$ is the change in surface temperature in response to a change in heat content, and $\lambda *$ is the climate feedback parameter (Schneider and Dickinson, 1974 [7]), also known as the climate sensitivity parameter, which denotes the rate at which the climate system returns the added forcing to space as infrared radiation or as reflected solar radiation (by changes in clouds, ice and snow, etc.). In essence, $\lambda *$ accounts for how feedbacks modify the surface temperature response to the forcing. In principle, $T'$ should account for changes in the temperature of the surface and the troposphere, and since the lapse rate is assumed to be known or is assumed to be a function of the surface temperature, $T'$ can be approximated by the surface temperature. For steady state, the solution yields

$$T' = f\, \lambda* \tag{1-3}$$

Recently, Pielke et al. [8] thoroughly discussed what a global average surface temperature does really mean. The authors, for instance, stated that, as a climate metric to diagnose climate system heat changes (i.e., ''global warming''), the surface temperature trend, especially if it includes the trend in nighttime temperature, is not the most suitable climate metric.

On the other hand, an excerpt from the Chapter 2 of the 4$^{th}$ report of the Working Group I to the Intergovernmental Panel on Climate Change (IPCC), "Climate Change 2007 – The Physical Science Basis" [9] reads:

> The definition of $RF$ from the TAR and earlier IPCC assessment reports is retained. Ramaswamy et al. (2001) define it as 'the change in net (down minus up) irradiance (solar plus longwave; in $W\,m^{-2}$) at the tropopause after allowing for stratospheric temperatures to readjust to radiative equilibrium, but with surface and tropospheric temperatures and state held fixed at the unperturbed values'. ….. Radiative forcing can be related through a linear relationship to the global mean equilibrium temperature change at the surface $(\Delta T_s)$: $\Delta T_s = \lambda * RF$, where $\lambda *$ is the climate sensitivity parameter (e.g., Ramaswamy et al., 2001 [10]).

Obviously, from both excerpts we may conclude that $f = RF$ and $T' = \Delta T_S$. Unfortunately, the notion "climate feedback parameter" used in the NRC report is incorrect. As already explained by Dickinson [5] and Kiehl [11], the reciprocal of the (global) feedback parameter $\lambda$ gives the change of global temperature for a given radiative perturbation. This reciprocal, $\lambda* = 1/\lambda$, can be called the (global) sensitivity parameter. The physical units of this sensitivity parameter are, therefore, $m^2\,K/W$ as required by equation 1-1 of the NRC report. Ramanathan et al. [12] used the formula $\Delta T_S = RF/\lambda$, where they called $\lambda$ the climate feedback parameter. The authors also stated that $\lambda$ estimated by the hierarchy of simple and sophisticated climate models lies in the range of $1 < \lambda < 4$ W m$^{-2}$ K$^{-1}$. In the case of a blackbody Dickinson [5] and Ramanathan et al. [12] estimated $\lambda$ for a radiating temperature of about 255 K by $\lambda_e = 4 \sigma T_e^3 \approx 3.7$ W m$^{-2}$ K$^{-1}$. Note that $T_e$ is the temperature inferred from the so-called planetary radiative equilibrium of the Earth in the absence of the atmosphere (see formula (3.20)). Here, it is called the radiative equilibrium temperature. Furthermore, $\lambda$ is also used to define the so-called gain factor by $g = 1 - \lambda/\lambda_e$ [4].

The simple climate model (1.1) or (1.3) requires not only a physical clarification, but also recognition of its uncertainties. Scrutinizing the uncertainties involved is indispensable when empirical quantities are used to quantify physical processes. Thus, in the following Eq. (1.1) and its solution are discussed, where especially the effect of the linearization of the power law of Stefan [13] and Boltzmann [14] and the role of the feedback parameter are debated (see section 2). In section 3 the common quantification of the greenhouse effect related to the radiative equilibrium temperature is scrutinized and the uncertainties involved are assessed using Gaussian error principles. In section 4 the correct solution of the global energy balance model is presented and its results are compared with the radiative equilibrium temperature. In section 5 we discuss the surface temperatures obtained when the down-welling infrared radiation is parameterized using either Ångström-type or Brunt-type formulae. A Dines-type two-layer energy balance model for the Earth-atmosphere system that contains the planetary radiation balance for an Earth in the absence of an atmosphere as an asymptotic solution is presented in section 6 and its results are briefly, but thoroughly discussed.

## 2. The global energy balance model of Schneider and Mass

### 2.1 Basic considerations

The global energy balance model of Schneider and Mass [2] for an aqua planet reads (see Appendix A)

$$R \frac{dT_s}{dt} = \left(1 - \alpha_E\right) \frac{S}{4} + F_{IR\downarrow} - F_{IR\uparrow}\left(T_s\right) \ , \tag{2.1}$$

where $S$ is the solar constant, $\alpha_E$ is the planetary integral albedo of the Earth, and $R$ is called the planetary inertia coefficient [15]. The factor 4 is called the geometry factor; it is based on the planetary radiation balance (see section 3). Usually, a value for the solar constant close to $S \cong 1367 \text{ W m}^{-2}$ is recommended (e.g., [16-18]), but the value provided by recent satellite observations using TIM (Total Irradiance Monitoring; satellite launched in 2003) is close to $S \cong 1361 \text{ W m}^{-2}$ (see Figure 1). This value is $4 \text{ W m}^{-2}$ higher than the 30-year mean reported by the Smithsonian Institution [19], but completely agrees with that of Laue and Drummond [20] which is based on direct observations at an altitude of 82 km.

The quantity $F_{IR\downarrow}$ is the flux density (hereafter simply called a flux) of down-welling infrared (IR) radiation. Estimates for this IR flux may be obtained using either Ångström-type or Brunt-type formulae that are related to the power law of Stefan [13] and Boltzmann [14] using near surface observations of temperature, humidity, and normalized cloud cover (see section 5). Furthermore, $F_{IR\uparrow}(T_s)$ is the terrestrial radiation emitted by the Earth's surface. This IR flux is calculated using the Stefan-Boltzmann law, too. It is given by

$$F_{IR\uparrow}\left(T_s\right) = \varepsilon_E \ \sigma \ T_s^4 \ , \tag{2.2}$$

where $\varepsilon_E \leq 1$ is the planetary integral emissivity, and $\sigma = 5.67 \cdot 10^{-8} \text{ W m}^{-2} \text{ K}^{-4}$ is Stefan's constant. Schneider and Mass [2], however, did not use this power law, neither for $F_{IR\downarrow}$ nor for $F_{IR\uparrow}(T_s)$. Instead, Budyko's [21, 22] empirical formula by

$$\Delta F_{IR} = F_{IR\uparrow}\left(T_s\right) - F_{IR\downarrow} = a + b\left(T_s - T_r\right) - \left\{a_1 + b_1\left(T_s - T_r\right)\right\} n \tag{2.3}$$

was considered with the empirical coefficients $a = 226.0 \text{ W m}^{-2}$, $b = 2.26 \text{ W m}^{-2} \text{ K}^{-1}$, $a_1 = 48.4 \text{ W m}^{-2}$, $b_1 = 1.61 \text{ W m}^{-2} \text{ K}^{-1}$, the reference temperature $T_r = 273.15 \text{ K}$, and the normalized cloud cover $n$. As already mentioned by Budyko [22], this formula completely agrees with that of Manabe and Wetherald [23],

$$\Delta F_{IR} = a + b(T_s - T_r) - a_2 \, n \tag{2.4}$$

with $a_2 = 25.8 \text{ W m}^{-2}$, if clear-sky conditions (i.e., in the absence of clouds) are assumed. This means that the empirical formulae (2.3) and (2.4) imply all radiative effects in the infrared range, i.e., even the absorption and emission of infrared radiation by the so-called greenhouse gases having either natural or anthropogenic origin. Furthermore, even though the planetary albedo also implies the effect of clouds in the solar range, Schneider and Mass [2] did not discuss the cloud effects in the infrared range. Moreover, these authors lowered Budyko's values to $a = 201.5 \text{ W m}^{-2}$ and $b = 1.45 \text{ W m}^{-2} \text{ K}^{-1}$. Based on the linearization of Sellers' [24] radiation formula, North [25] and Kiehl [11] recommended $a = 211.2 \text{ W m}^{-2}$ and $b = 1.55 \text{ W m}^{-2} \text{ K}^{-1}$. Even though Eq. (2.1) is considered as a global energy balance model, the fact is that this model is only based on a radiation balance because the exchange of energy between the Earth's skin and the atmosphere by the fluxes of sensible and latent heat is completely ignored. Unfortunately, this radiation balance is rather imperfect because the absorption of solar radiation by atmospheric constituents is ignored, too.

Nevertheless, inserting formula (2.3) into Eq. (2.1) and considering clear-sky conditions provides

$$R \frac{dT_s}{dt} = Q - \lambda \, T_s \tag{2.5}$$

with

$$Q = (1 - \alpha_E)\frac{S}{4} - a + b\, T_r \tag{2.6}$$

and $\lambda = b$. If we interpret $T_s$ as a generalized coordinate and $dT_s/dt$ as the corresponding generalized velocity and assume that $Q$ is independent of time, we may transfer Eq. (2.5) into the phase space, where it is considered as a one-dimensional model because it has only one degree of freedom [26]. As the feedback parameter, $\lambda$, is positive, the solution of Eq. (2.5) tends to an attractor given by $dT_s/dt = 0$, the condition of the fixed point.

Apparently, there is a notable difference between Eq. (1.1) used by various authors (e.g., [1, 3, 5]) and Eq. (2.5), namely the difference between $C = \rho\, c$ and $R$, where $\rho$ is, again, the density of the material under study, and $c$ is the corresponding specific heat. As shown in the Appendix A, this difference can be expressed by

$$R = C\, \vartheta \;, \tag{2.7}$$

i.e., the planetary inertia coefficient is equal to the heat capacity of the system times a length scale, here the thickness $\vartheta$ of the water layer of an aqua planet as considered by Schneider and Mass [2]. Note that the physical units of the heat capacity are $J\, m^{-3}\, K^{-1}$, while those of the planetary inertia coefficient are $J\, m^{-2}\, K^{-1}$ as requested by equations (2.5) and 1-1 of the NRC report.

The linearization of an IR term as reflected by formula (2.3) can simply be explained, for instance, in the case of the emitted IR radiation by expressing the surface temperature as $T_s = T_r + \Delta T_s$. Thus, we may write

$$F_{IR\uparrow}(T_s) = \varepsilon_E\, \sigma\, T_s^{\,4} = \varepsilon_E\, \sigma\, T_r^{\,4}\left(1 + \frac{\Delta T_s}{T_r}\right)^{\!4} . \tag{2.8}$$

Since usually $|\Delta T_s/T_r| \ll 1$, we may write $(1 + \Delta T_s/T_r)^4 \approx 1 + 4\,\Delta T_s/T_r$. This approximation leads to

$$F_{IR\uparrow}(T_s) = 4\,\varepsilon_E\,\sigma\,T_r^3\,T_s - 3\,\varepsilon_E\,\sigma\,T_r^4 \qquad (2.9)$$

Inserting this equation into Eq. (2.1) yields (e.g., [4])

$$R\frac{dT_s}{dt} = (1-\alpha_E)\frac{S}{4} + 3\,\varepsilon_E\,\sigma\,T_r^4 - 4\,\varepsilon_E\,\sigma\,T_r^3\,T_s \quad . \qquad (2.10)$$

for the Earth in the absence of its atmosphere. Now, with respect to Eq. (2.5) the quantities Q and $\lambda$ may be identified by

$$Q = (1-\alpha_e)\frac{S}{4} + 3\,\varepsilon_E\,\sigma\,T_r^4 \qquad (2.11)$$

and

$$\lambda = 4\,\varepsilon_E\,\sigma\,T_r^3 \quad , \qquad (2.12)$$

respectively. Thus, in such a case Q and $\lambda$ depend on the reference temperature $T_r$ (see Figure 2). Obviously, for $T_r \geq 260\,K$ the feedback parameter $\lambda$ exceeds $4\,W\,m^{-2}\,K^{-1}$. Consequently, the statement of Ramanathan et al. [12] that $\lambda$ lies in the range of $1 < \lambda < 4\,W\,m^{-2}\,K^{-1}$ is only valid for reference temperatures less than $260\,K$ when the Earth in the absence of its atmosphere is considered.

    Since Schneider and Mass [2] expressed the solar constant in Eq. (2.6) as a function of time due to the time-dependent influence of solar activity (characterized by sunspot numbers) and atmospheric dust (provided, for instance, by volcano eruptions), they presented results from numerical solutions of Eq. (2.5). If we, however, assume that $\alpha_E$, $\varepsilon_E$, and S, and hence, Q are independent of time as assumed by various authors (e.g., [1, 3, 4]), the exact solution of Eq. (2.5) is given by

$$T_s(t) = T_{s0} \exp\left(-\frac{\lambda}{R} t\right) + \frac{Q}{\lambda}\left(1 - \exp\left(-\frac{\lambda}{R} t\right)\right) \ , \tag{2.13}$$

where $T_{s0} = T_s(t=0)$ is the initial temperature. If time tends to infinity, we will obtain: $T_s(\infty) = T_s(t \to \infty) = Q/\lambda$ also called the radiative equilibrium temperature [1] or the fixed point temperature that is based on the condition: $dT_s/dt = 0$. Thus, for any initial temperature $T_s(t)$ tends to $T_s(\infty)$, i.e., as aforementioned, the radiative equilibrium temperature is an attractor. Defining the so-called e-folding time of the response, $\tau = R/\lambda$, yields finally

$$T_s(t) = T_{s0} \exp\left(-\frac{t}{\tau}\right) + T(\infty)\left(1 - \exp\left(-\frac{t}{\tau}\right)\right) \ . \tag{2.14}$$

Note that this equation completely agrees with formula (21) of Hansen et al. [4].

Several authors assumed that the initial temperature $T_{s0}$ is equal to zero (e.g., [1, 3]) so that Eq. (2.14) results in

$$T_s(t) = T(\infty)\left(1 - \exp\left(-\frac{t}{\tau}\right)\right) \ . \tag{2.15}$$

Since $T_s(\infty)$ is an attractor, this assumption does not affect the result for the radiative equilibrium temperature.

*2.2  The inclusion of the absorption of solar radiation and the exchange of sensible and latent heat*

As mentioned before, the global energy balance model of Schneider and Mass [2] does not include the absorption of solar radiation by atmospheric constituents and the exchange of sensible and latent heat between the water layer of the aqua planet and the atmosphere. If the atmospheric absorption, $A_a S$, and the (vertical components of the) fluxes of sensible heat, $H$, and latent heat, $E = L_v(T_s) Q$, at the surface of the water layer are inserted into Eq. (2.1), where $A_a$ is the integral absorptivity with respect to the range of solar radiation, $L_v(T_s)$ is the specific heat of phase transition (e.g., vaporization, sublimation), considered as dependent on the surface temperature, $T_s$, and $Q$ is the water vapor flux, we will obtain (see Appendix)

$$R \frac{dT_s}{dt} = \left(1 - \alpha_E - A_a\right) \frac{S}{4} - H - E - \Delta F_{IR} \quad . \tag{2.16}$$

Using Eq. (2.3) for clear-sky conditions leads to Eq. (2.5) again, but the forcing term (2.6) becomes

$$Q = \left(1 - \alpha_E - A_a\right) \frac{S}{4} - H - E - a + b\, T_r \quad . \tag{2.17}$$

Recent estimates for the fluxes of sensible and latent heat result in $H \cong 17\ \text{W m}^{-2}$ and $E \cong 80\ \text{W m}^{-2}$ if the global Bowen ratio of $Bo = 0.21$ is considered [27] (see also Figure 3). These values, however, differ from those of Schneider [15] and Liou [16] showing that even in this case a notably degree of uncertainty exists. Based on the accuracy of direct measurements of turbulent fluxes and related parameterization schemes [28] we have to assume that an uncertainty of more than 15 percent in both fluxes exists. Following Liou [16] a value of $A_a = 0.23$ is chosen to predict the atmospheric absorption of solar radiation. This value completely agrees with that of Trenberth et al. [27], but is slightly lower than that of Schneider [15]. Taking $\alpha_E = 0.30$ and $A_a = 0.23$ into account, the solar radiation absorbed by the Earth's skin results in $\left(1 - \alpha_E - A_a\right) S/4 = 160.6\ \text{W m}^{-2}$ (see Figure 3).

*2.3   The effects of clouds in the infrared range*

Obviously, formulae (2.3) and (2.4) independently obtained by Budyko [21, 22] and Manabe and Wetherald [23] coincide in the case of a clear sky. Since the planetary albedo also implies the albedo of clouds, it is indispensable to consider the effect of clouds in the infrared range. Thus, in the case of a cloudy sky the quantities $Q$ and $\lambda$ occurring in Eq. (2.5) read

$$Q = \begin{cases} \left(1 - \alpha_E - A_a\right)\dfrac{S}{4} - H - E - a + a_1 n + \left(b - b_1 n\right) T_r & \text{if [21, 22]} \\ \left(1 - \alpha_E - A_a\right)\dfrac{S}{4} - H - E - a + a_2 n + b\, T_r & \text{if [23]} \end{cases} \qquad (2.18)$$

and

$$\lambda = \begin{cases} b\left(1 - \dfrac{b_1}{b} n\right) & \text{if [21, 22]} \\ b & \text{if [23]} \end{cases} \qquad (2.19)$$

Assuming a stationary state (i.e., $dT_s/dt = 0$) yields for Budyko's formula (see also Eq. (2.8) by Budyko [22])

$$T_s(\infty) = \dfrac{Q}{\lambda} = T_r + \dfrac{1}{b\left(1 - \dfrac{b_1}{b} n\right)}\left\{\left(1 - \alpha_E - A_a\right)\dfrac{S}{4} - H - E - a\left(1 - \dfrac{a_1}{a} n\right)\right\} \qquad (2.20)$$

and for the formula of Manabe and Wetherald [23]

$$T_s(\infty) = \dfrac{Q}{\lambda} = T_r + \dfrac{1}{b}\left\{\left(1 - \alpha_E - A_a\right)\dfrac{S}{4} - H - E - a\left(1 - \dfrac{a_2}{a} n\right)\right\} \; . \qquad (2.21)$$

*2.4    The anthropogenic radiative forcing*

According to Forster et al. [9], the RF represents the stratospherically adjusted radiative flux change evaluated at the tropopause, as defined by Ramaswamy et al. [10]. Positive RF-values lead to a global mean surface warming and negative RF-values to a global mean surface cooling.

If we insert the radiative forcing, RF, due to anthropogenic activities into Eq. (2.17), relevant for clear-sky conditions, we will obtain

$$Q = \left(1 - \alpha_E - A_a\right)\frac{S}{4} - H - E - a + b\,T_r + RF \quad . \tag{2.22}$$

As illustrated in Figure 4, the globally averaged net anthropogenic radiative forcing in 2005 corresponds to $RF = 1.6\,(0.6 \text{ to } 2.4)\,W\,m^{-2}$ relative to pre-industrial conditions defined at 1750 [9]. This value may be considered as a perturbation to the natural system [11]. However, we have to recognize that (1) a portion of this RF is already covered by empirical constants $a$ and $b$ (depending on the period of observation of infrared radiation from which $a$ and $b$ are deduced) completely ignored by the Working Group I of the IPCC [9], and (2) the value of RF is much smaller than the uncertainty with which the atmospheric absorption of solar radiation, the coefficient $a$, and the fluxes of sensible and latent heat are fraught.

*2.5  Estimates*

First, Eqs. (2.6) and (2.17) were alternatively used to estimate the radiative equilibrium temperature $T_s(\infty) = Q/\lambda$, where $\alpha_E = 0.30$. In the case of Eq. (2.17) $A_a = 0.23$ was assumed in an additional sensitivity study. The estimates are listed in Table 1. For the purpose of comparison: the globally averaged near-surface temperature that is based on routine observations using the stations of the global meteorological network and weather satellites amounts to $\langle T_{ns} \rangle \approx 288\,K$. Thus, ignoring the fluxes of sensible and latent heat (Eq. (2.6)) leads to $T_s(\infty)$ results that are ranging from 278.9 K to 299 K, i.e., these temperature values are much higher than the temperature $T_e \approx 255\,K$ inferred from the planetary radiative equilibrium of the Earth in the absence of an atmosphere (see Eq. (3.20)) even though they notably disagree with $\langle T_{ns} \rangle$. However, they become appreciably lower than $T_e$ when $H$ and $E$ are included (see Eq. (2.17)). They become still lower if the absorption of solar radiation by atmospheric constituents is included additionally (see Table 1). Since the special case of the planetary radiative equilibrium of the Earth in the absence of an atmosphere is an asymptotic solution of either formula (2.20) or formula (2.21), the global energy balance model of Schneider and Mass [2] does not support the existence of the so-called atmospheric greenhouse effect, as defined, for instance, by the American Meteorological Society (see section 3).

The results provided by formulae (2.20) and (2.21) for different values of the normalized cloud cover are illustrated in Figure 5. The effect of the cloud cover on the surface temperature is notable. Budyko's [22] empirical formula predicts a decrease in the surface temperature when the cloud cover rises. The opposite is true in the case of the empirical formula of Manabe and Wetherald [23].

These facts document that the entire concept of the global energy balance model of Schneider and Mass [2] expressed by Eq. (2.1) is highly inconsistent because the fluxes of sensible and latent heat and the absorption of solar radiation by atmospheric constituents are not negligible. As already mentioned, this global energy balance model is only a global radiation balance model that does not exist in the case of the real Earth-atmosphere system. Consequently, all of these values for the empirical constants especially $a$ and $b$ are unreasonable.

The effect of the net anthropogenic radiative forcing in 2005 of $RF = 1.6 \, W \, m^{-2}$, on average, was also estimated using $RF/\lambda$. These estimates are listed in Table 1, too. Obviously, this net anthropogenic radiative forcing causes a slight increase in the surface temperature ranging from $\Delta T_s = 0.71 \, K$ to $\Delta T_s = 1.11 \, K$. Also these $\Delta T_s$-values underline that the effect of $RF$, the perturbation to the natural system, is much smaller than the uncertainty involved in the solution of Eq. (2.5) using formulae (2.6) and (2.17) as well as the empirical constants $a$ and $b$. Furthermore, rearranging Eq. (2.20) (or Eq. (2.21)) yields

$$\lambda = \frac{1}{T_s(\infty) - T_r} \left\{ \left(1 - \alpha_E - A_a\right) \frac{S}{4} - H - E - a\left(1 - \frac{a_1}{a} n\right) \right\} , \qquad (2.23)$$

i.e., for $a\left(1 - \frac{a_1}{a} n\right) \geq \left(1 - \alpha_E - A_a\right) \frac{S}{4} - H - E$, as expected on the basis of the empirical values mentioned above, and any positive temperature difference, $T_s(\infty) - T_r$, the feedback parameter would be either equal to zero or negative. Since the effect of the anthropogenic radiative forcing is given by $RF/\lambda$, the latter would cause a decrease of the surface temperature.

This example shows how useless the feedback equation (2.5) is. Moreover, either Eq. (2.14) or formula (2.15) provide radiative equilibrium temperatures for the Earth in the absence of its atmosphere, which notably differ from the true solution of Eq. (2.1), when Eq. (2.2) is considered. This can be demonstrated the best on the basis of the Earth's radiation balance in the case of the absence of its atmosphere.

## 3. The planetary radiation balance in the absence of an atmosphere and the greenhouse effect

The so-called greenhouse effect of the atmosphere is commonly explained as followed (see Glossary of Meteorology, American Meteorological Society, http://amsglossary.allenpress.com/glossary/search?id=greenhouse-effect1):

> "The heating effect exerted by the atmosphere upon the Earth because certain trace gases in the atmosphere (water vapor, carbon dioxide, etc.) absorb and reemit infrared radiation.
>
> Most of the sunlight incident on the Earth is transmitted through the atmosphere and absorbed at the Earth's surface. The surface tries to maintain energy balance in part by emitting its own radiation, which is primarily at the infrared wavelengths characteristic of the Earth's temperature. Most of the heat radiated by the surface is absorbed by trace gases in the overlying atmosphere and reemitted in all directions. The component that is radiated downward warms the Earth's surface more than would occur if only the direct sunlight were absorbed. The magnitude of this enhanced warming is the greenhouse effect. Earth's annual mean surface temperature of 15°C is 33°C higher as a result of the greenhouse effect than the mean temperature resulting from radiative equilibrium of a blackbody at the Earth's mean distance from the sun. The term "greenhouse effect" is something of a misnomer. It is an analogy to the trapping of heat by the glass panes of a greenhouse, which let sunlight in. In the atmosphere, however, heat is trapped radiatively, while in an actual greenhouse, heat is mechanically prevented from escaping (via convection) by the glass enclosure."

According to this explanation we may carry out the following "thought experiment" of a planetary radiation equilibrium also called a zero-dimensional model, where we assume an Earth without an atmosphere. A consequence of this assumption is that (a) absorption of solar and terrestrial (infrared) radiation, (b) scattering of solar radiation by molecules and particulate matter, (c) radiative emission of energy in the infrared range, (d) convection and advection of heat, and (e) phase transition processes related to the formation and depletion of clouds can be ignored. This thought experiment is carried in detail to point out that several assumptions and simplifications are necessary to obtain the formula of the planetary radiative equilibrium. This formula serves to predict the radiative equilibrium temperature $T_e$.

### 3.1 Solar radiation

The total flux (also called the radiant power) of solar radiation, $E_{S\downarrow}$, reaching the surface of an Earth without an atmosphere is given by

$$E_{S\downarrow} = \int_A \mathbf{F} \cdot \mathbf{n}(\mathbf{r}) \, dA(\mathbf{r}) \quad . \tag{3.1}$$

Here, $\mathbf{F}$ is the flux of solar radiation also called the solar irradiance and F is its magnitude, A is the radiation-exposed surface of the Earth, $dA(\mathbf{r})$ the corresponding differential surface element, and $\mathbf{n}(\mathbf{r})$ is the inward pointing unit vector perpendicular to $dA(\mathbf{r})$, where the direction of $\mathbf{n}(\mathbf{r})$ is chosen in such a sense that $\mathbf{F} \cdot \mathbf{n}(\mathbf{r}) \geq 0$ is counted positive. This unit vector is also called the unit normal. Its origin is the tip of the position vector $\mathbf{r}$ with which the location of $dA(\mathbf{r})$ on the Earth's surface is described. The angle between $\mathbf{F}$ and $\mathbf{n}(\mathbf{r})$ is the local zenith angle of the Sun's center, $\Theta_0(\mathbf{r})$. The scalar product $\mathbf{F} \cdot \mathbf{n}(\mathbf{r})$ that describes the solar radiation reaching the horizontal surface element, is given by $\mathbf{F} \cdot \mathbf{n}(\mathbf{r}) = F \cos \Theta_0(\mathbf{r})$.

From now on we consider the Earth as a sphere; and its center as the origin of the position vector so that $\mathbf{r} = \mathbf{r}_E$, where $r_E = 6371\,\mathrm{km}$ is its magnitude. Thus, $\mathbf{n}(\mathbf{r}_E)$ and $\mathbf{r}_E$ are collinear vectors having opposite directions. Furthermore, $A = r_E^2 \, \Omega$ with $\Omega = 2\pi$ is the solid angle of a half sphere, and $dA = r_E^2 \, d\Omega$, where $d\Omega = \sin\theta \, d\theta \, d\phi$ is the differential solid angle (see Figure 6). The quantities $\theta$ and $\phi$ are the zenith and azimuthal angles, respectively, of a spherical coordinate frame (see Figure 6). Thus, the radiant power can be written as

$$E_{S\downarrow} = r_E^2 \int_\Omega F \cos \Theta_0 \, d\Omega = r_E^2 \int_0^{2\pi} \int_0^{\pi/2} F \cos \Theta_0 \, \sin\theta \, d\theta \, d\phi \quad . \tag{3.2}$$

Since, however, a portion of $F \cos \Theta_0(\mathbf{r}_E)$ is not absorbed by the skin of the Earth's surface because it is diffusely reflected we have to insert the local albedo, $\alpha(\Theta_0, \theta, \phi)$, into this equation. Note that the local albedo not only depends on $\theta$ and $\phi$, but also on $\Theta_0(\mathbf{r}_E)$. The total flux of solar radiation that is absorbed by the Earth's skin, $F_{S\downarrow}$, is then given by

$$F_{S\downarrow} = r_E^2 \int_\Omega F(1 - \alpha(\Theta_0, \theta, \phi)) \cos \Theta_0 \, d\Omega = r_E^2 \int_0^{2\pi} \int_0^{\pi/2} F(1 - \alpha(\Theta_0, \theta, \phi)) \cos \Theta_0 \, \sin\theta \, d\theta \, d\phi \tag{3.3}$$

The solar irradiance reaching the Earth's surface is given by

$$F = \left(\frac{r_S}{d}\right)^2 F_S \quad, \tag{3.4}$$

where $r_S = 6.96 \cdot 10^6$ km is the visible radius of the Sun, d is the actual distance between the Sun's center and the orbit of the Earth, and $F_S$ denotes the solar emittance [16]. Formula (3.4) is based on the fact that the radiant power ($= 4 \pi r_S^2 F_S$) of the Sun is kept constant when the solar radiation is propagating through the space because of energy conservation principles in the absence of an intervening medium [16, 30, 31]. For the mean distance (1 Astronomic Unit = AU) of $d_0 = 1.496 \cdot 10^8$ km (e.g., [32, 33]), F becomes the solar constant S so that we may write (e.g., [16, 32, 34])

$$F = \left(\frac{d_0}{d}\right)^2 S \quad . \tag{3.5}$$

The quantity $(d_0/d)^2$ is denoted here as the orbital effect. Since $(d_0/d)^2$ does not vary more than 3.5 percent [16, 34], this orbital effect is usually ignored. Note, however, that the temperature difference between perihelion and aphelion amounts to $4.2\,\text{K}$. This temperature difference is much larger than the increase of the globally averaged near-surface temperature during the last 160 years (see the temperature anomaly with respect to the period 1961-1990 illustrated in Figure 7).

The local zenith angle of the Sun's center can be determined using the rules of spherical trigonometry. In doing so, one obtains

$$\cos \Theta_0 = \sin \varphi \sin \delta + \cos \varphi \cos \delta \cos h \quad , \tag{3.6}$$

where $\varphi$ is the latitude, $\delta$ is the solar declination angle, and h is the hour angle from the local meridian (e.g., [16, 32, 34]). The solar declination angle can be determined using $\sin \delta = \sin \beta \sin \gamma$, where $\beta = 23°\ 27'$ is the oblique angle of the Earth's axis, and $\gamma$ is the true longitude of the Earth counted counterclockwise from the vernal equinox (e.g., [16, 32]). The latitude is related to the zenith angle by $\varphi = \pi/2 - \theta$ so that formula (3.6) may be written as $\cos \Theta_0 = \cos \theta \sin \delta + \sin \theta \cos \delta \cos h$. Note that $\theta$ is ranging from zero to $\pi$, $\delta$ from $23°\ 27'\,\text{S}$ (Tropic of Capricorn) to $23° 27'$ N (Tropic of Cancer), and h from $-H$ to $H$, where $H$ represents the half-day, i.e., from sunrise to solar noon or solar noon to sunset. It can be deduced from Eq. (3.6) by setting $\Theta_0 = \pi/2$ (invalid at the poles) leading to $\cos H = -\tan\varphi \tan\delta$ (e.g., [16, 32, 34]).

By considering these issues Eq. (3.3) may be written as

$$F_{S\downarrow} = r_E^2 \left(\frac{d_0}{d}\right)^2 S \left\{ \int_0^{2\pi}\int_0^{\pi/2} \cos\Theta_0 \sin\theta \, d\theta \, d\phi - \int_0^{2\pi}\int_0^{\pi/2} \alpha(\Theta_0, \theta, \phi) \cos\Theta_0 \sin\theta \, d\theta \, d\phi \right\} \quad . \tag{3.7}$$

The equation cannot generally be solved in an analytical manner. Two special solutions, however, are possible.

If we assume, for instance, that $\delta = \pi/2$ we will obtain $\cos\Theta_0 = \cos\theta$, i.e., the rotation axis of the Earth would always be parallel to the incoming solar radiation. Note that (1) the value of $\delta$ exceeds the Tropic of Cancer by far and (2) the rotation of the planet plays no role. Obviously, this choice of $\delta$ results in

$$\begin{aligned} F_{S\downarrow} &= r_E^2 \left(\frac{d_0}{d}\right)^2 S \left\{ \int_0^{2\pi}\int_0^{\pi/2} \cos\theta \sin\theta \, d\theta \, d\phi - \int_0^{2\pi}\int_0^{\pi/2} \alpha(\theta, \phi) \cos\theta \sin\theta \, d\theta \, d\phi \right\} \\ &= r_E^2 \left(\frac{d_0}{d}\right)^2 S \left\{ \pi - \int_0^{2\pi}\int_0^{\pi/2} \alpha(\theta, \phi) \cos\theta \sin\theta \, d\theta \, d\phi \right\} \end{aligned} \quad , \tag{3.8}$$

Considering $(d_0/d)^2 \approx 1$ and assuming a constant albedo for the entire planet, i.e., $\alpha(\theta, \phi) = \alpha_E$, yield [2, 4, 15-17, 31-33, 36]

$$F_{S\downarrow} = \pi r_E^2 S (1 - \alpha_E) \quad . \tag{3.9}$$

This formula can also be derived for a rotating Earth in the absence of its atmosphere when $\delta = 0$ is assumed. This assumption is only fulfilled biyearly, namely at the vernal equinox and the autumnal equinox. In doing so, we consider the solar insolation that is defined as the flux of solar radiation per unit of horizontal area for a given location [16]. Thus, the daily solar insolation absorbed by the Earth's skin is given by (e.g., [16, 30, 34, 37])

$$f_{S\downarrow} = \int_{t_r}^{t_s} \left(\frac{d_0}{d}\right)^2 S\left(1 - \alpha(\Theta_0, \theta, \phi)\right) \cos \Theta_0 \, dt \quad . \tag{3.10}$$

Here, t is time, where $t_r$ and $t_s$ correspond to sunrise and sunset, respectively. Assuming, again, a constant albedo for the entire planet, i.e., $\alpha(\Theta_0, \theta, \phi) = \alpha_E$ and recognizing that the variation of the actual distance, d, between the Sun and the Earth in one day can be neglected yield

$$\left.\begin{aligned} f_{S\downarrow} &= \left(\frac{d_0}{d}\right)^2 S\left(1 - \alpha_E\right) \int_{t_r}^{t_s} \cos \Theta_0 \, dt \\ &= \left(\frac{d_0}{d}\right)^2 S\left(1 - \alpha_E\right) \int_{t_r}^{t_s} \left(\cos \theta \sin \delta + \sin \theta \cos \delta \cos h\right) dt \end{aligned}\right\} . \tag{3.11}$$

Defining the angular velocity of the Earth by $\omega = dh/dt = 2\pi \text{ rad/day}$, Eq. (3.11) may be written as

$$\left.\begin{aligned} f_{S\downarrow} &= \left(\frac{d_0}{d}\right)^2 \frac{S}{\omega}\left(1 - \alpha_E\right) \int_{-H}^{H} \left(\cos \theta \sin \delta + \sin \theta \cos \delta \cos h\right) dh \\ &= \left(\frac{d_0}{d}\right)^2 \frac{S}{\pi}\left(1 - \alpha_E\right)\left(H \cos \theta \sin \delta + \sin \theta \cos \delta \sin H\right) \end{aligned}\right\} , \tag{3.12}$$

where H is, again, the half-day. This equation described the daily solar insolation absorbed at a given location. Thus, the total amount of solar radiation that is absorbed by the planet during one of its rotations, $F_{S\downarrow}$, is given by

$$F_{S\downarrow} = \int_\Omega r_E^{\,2}\, f_{S\downarrow}\, d\Omega$$

$$= r_E^{\,2} \left(\frac{d_0}{d}\right)^2 \frac{S}{\pi}(1-\alpha_E) \int_0^{2\pi}\!\int_0^{\pi} (H\cos\theta\sin\delta + \sin\theta\cos\delta\sin H)\sin\theta\, d\theta\, d\phi \quad . \tag{3.13}$$

If we assume that the oblique angle of the planet's axis is zero which means that the axis is perpendicular to the plane of the ecliptic, we will obtain $\delta = 0$. Thus, the half-day will amount to $H = \pi/2$, i.e., the length of the solar day is 12 hours. In doing so, Eq. (3.13) becomes

$$F_{S\downarrow} = r_E^{\,2} \left(\frac{d_0}{d}\right)^2 \frac{S}{\pi}(1-\alpha_E)\int_0^{2\pi}\!\int_0^{\pi}\sin^2\theta\, d\theta\, d\phi = \pi\, r_E^{\,2}\left(\frac{d_0}{d}\right)^2 S(1-\alpha_E) \quad . \tag{3.14}$$

As mentioned before, $(d_0/d)^2$ does not vary more than 3.5 percent. Thus, ignoring the orbital effect again leads to formula (3.9).

Even though some drastic assumptions and simplifications were introduced, formula (3.9) is customarily used to quantify the incoming solar radiation that is absorbed at the Earth's surface [2, 4, 15-17, 31-33, 36].

*3.2 Terrestrial radiation*

The total flux of infrared radiation emitted by the Earth's surface is given by

$$F_{IR\uparrow} = \int_A \mathbf{E}(\mathbf{r}_E)\cdot \mathbf{n}(\mathbf{r}_E)\, dA(\mathbf{r}_E) \quad . \tag{3.15}$$

Here, $\mathbf{E}(\mathbf{r}_E) = \varepsilon(\mathbf{r}_E)\sigma T^4(\mathbf{r}_E)\mathbf{e}_r$ is the flux density of infrared radiation at the location $\mathbf{r}_E$, and $E(\mathbf{r}_E) = \varepsilon(\mathbf{r}_E)\sigma T^4(\mathbf{r}_E)$ is its magnitude. The emission of radiant energy by a small surface element into the adjacent half space is considered as isotropic, as required by the derivation of the power law of Stefan [13] and Boltzmann [14]. The unit vector $\mathbf{e}_r = \mathbf{r}_E/|\mathbf{r}_E|$ points to the zenith of this adjacent half space. Note that $\mathbf{e}_r$ and $\mathbf{n}(\mathbf{r}_E)$ are collinear unit vectors having

opposite directions. Thus, we have $\mathbf{e}_r \cdot \mathbf{n}(\mathbf{r}_E) = -1$. This only means that $F_{S\downarrow}$ is counted positive, and $F_{IR\uparrow}$ is counted negative. In the case of the planetary radiative equilibrium this convention results in

$$F_{S\downarrow} - F_{IR\uparrow} = 0 \quad \Leftrightarrow \quad F_{S\downarrow} = F_{IR\uparrow} \quad . \tag{3.16}$$

The radiant power in the infrared range is then given by

$$F_{IR\uparrow} = r_E^2 \int_\Omega E(\theta, \phi) d\Omega = r_E^2 \int_0^{2\pi} \int_0^\pi E(\theta, \phi) \sin\theta \, d\theta \, d\phi \quad , \tag{3.17}$$

where now $\Omega = 4\pi$ is the solid angle for the entire planet, and $E(\theta, \phi) = \varepsilon(\theta, \phi) \sigma T^4(\theta, \phi)$. The emissivity $\varepsilon(\theta, \phi)$ and the surface temperature $T(\theta, \phi)$ depend on both $\theta$ and $\phi$. Usually, it is assumed that $E(\theta, \phi)$ is uniformly distributed and may be substituted by $E_E = \varepsilon_E \sigma T_e^4$, where, again, $\varepsilon_E$ denotes the planetary emissivity. We obtain

$$F_{IR\uparrow} = r_E^2 \varepsilon_E \sigma T_e^4 \int_0^{2\pi} \int_0^\pi \sin\theta \, d\theta \, d\phi = 4\pi r_E^2 \varepsilon_E \sigma T_e^4 \quad . \tag{3.18}$$

This equation is customarily used to quantify the radiant power in the infrared range [2, 4, 15-17, 31-33, 36]. Note that the drastic assumption that $E(\theta, \phi)$ is uniformly distributed is not in agreement with our experience because even in the case of the real Earth-atmosphere system there is a variation of the Earth's surface temperature, at least, from the equator to the poles. Our Moon, for instance, nearly satisfies the requirement of a planet without an atmosphere; and formula (3.9) may similarly be applied to determine the solar radiation absorbed by its skin using a planetary albedo of $\alpha_M = 0.12$. It is well known that the Moon has no uniform temperature, and, hence, no uniform distribution of $E(\theta, \phi)$. There is not only a strong variation of the Moon's surface temperature from the lunar day to the lunar night, but also from the Moon equator to its poles (e.g., [38-40]). Gerlich and Tscheuschner [41] disputed this drastic assumption in detail.

## 3.3 The temperature of the radiative equilibrium

The planetary radiative equilibrium, Eq. (3.16), leads to (e.g., [2, 4, 15-17, 31-33, 36, 42])

$$S(1-\alpha_E) = 4\varepsilon_E \sigma T_e^4 \tag{3.19}$$

and in a further step to

$$T_e = \left\{ \frac{(1-\alpha_E)S}{4\varepsilon_E \sigma} \right\}^{\frac{1}{4}}, \tag{3.20}$$

where $T_e$ is, again, the radiative equilibrium temperature, i.e., the temperature inferred from the planetary radiative equilibrium. Note, however, that (1) the assumption that $E(\theta, \phi)$ is uniformly distributed is entirely unrealistic as debated before, and (2) the transfer of heat into the layer of the thickness $\vartheta$ beneath the Earth's surface is ignored. The latter is in contradiction to the formulation of Schneider and Mass [2] for the aqua planet. However, in the case of a planetary radiative equilibrium the left-hand side of Eq. (2.1) becomes equal to zero because $dT_s/dt \to 0$ if $T_s(t) \to T_s(\infty)$ so that this contradiction has no effect under such conditions.

Assuming that the Earth acts like a black body ($\varepsilon_E = 1$) and choosing $\alpha_E = 0.30$ as obtained for the Earth-atmosphere system from satellite observations (e.g., [31, 33]) yield $T_e \cong 254.86\,\text{K}$. This value is well known and the difference between the globally averaged near-surface temperature of nearly $\langle T_{ns} \rangle \approx 288\,\text{K}$ and this value of $T_e$ is customarily used to quantify the so-called greenhouse effect of about $\langle T_{ns} \rangle - T_e \approx 33\,\text{K}$ (e.g., [4, 22, 31, 33]). However, as illustrated in Figure 8, Eq. (2.14) provides $T_s(\infty) = 256.62\,\text{K}$, when a reference temperature $T_r = 273.15\,\text{K}$ is chosen as suggested by Budyko [21, 22] and Manabe and Wetherald [23]. This means that the radiative equilibrium temperature provided by Eq. (2.14) is 1.76 K higher than the true radiative equilibrium temperature, or in other words: Eq. (2.14) provides a "temperature increase" which is based on an inaccurate solution of Eq. (2.1) due to the linearization of the power law of Stefan and Boltzmann (see formulae (2.8) and (2.9)). This systematic error of a temperature increase of $1.76\,\text{K}$ K corresponds to $6.7\,\text{W m}^{-2}$ which is four times larger than the net anthropogenic radiative forcing of $\text{RF} = 1.6\,\text{W m}^{-2}$, on average, in 2005 (see Figure 4). Only reference temperatures in close vicinity to the radiative equilibrium temperature provide appreciably better results (see Figure 8). Note that Budyko [22] also calculated an radiative equilibrium temperature by assuming $S \cong 1360\,\text{W m}^{-2}$, $\varepsilon_E = 0.95$ and $\alpha_E \cong 0.33$. He obtained $T_e \cong 255\,\text{K}$. It seems that the assumptions have been made in such a sense that the amount of the greenhouse effect of about $33\,\text{K}$ can be kept. He noted, however, that in the absence of an atmosphere the planetary albedo cannot be equal to the actual value of $\alpha_E \cong 0.33$ (today $\alpha_E = 0.30$). He assumed that prior to the origin of the atmosphere, the Earth's albedo was lower and probably differed very little from the Moon's albedo, which is equal to $\alpha_M \cong 0.07$ (today $\alpha_M \cong 0.12$).

It is well known that the Earth is a gray body emitter, rather than a blackbody emitter. Therefore, we have to consider a planetary emissivity less than unity. In doing so, formula (3.20) should be expressed for the purpose of convenience by

$$T_e = \varepsilon_E^{-\frac{1}{4}} \underbrace{\left( \frac{(1 - \alpha_E) S}{4 \sigma} \right)^{\frac{1}{4}}}_{T_{b,e}} = \varepsilon_E^{-\frac{1}{4}} T_{b,e} \quad , \tag{3.21}$$

where $T_{b,e}$ is the blackbody equilibrium temperature. As illustrated in Figure 9, if we assume, for instance, $\varepsilon_E = 0.90$ we will obtain $T_e \approx 261.8\,\text{K}$ and, hence, $\Delta T = \langle T_{ns} \rangle - T_e \approx 26.2\,\text{K}$. For $\varepsilon_E = 0.75$ we would obtain $T_e \approx 274\,\text{K}$ and, hence, $\Delta T \approx 14\,\text{K}$. Assuming $\Delta T = 0$ yields $\varepsilon_E \cong 0.614$, i.e., to guarantee that a greenhouse effect can be identified the planetary emissivity has to satisfy the requirement $\varepsilon_E > 0.614$ (see Figure 9). Our estimates are not surprising because a decreasing planetary emissivity would lead to a decreasing emission of radiative energy, which means that the cooling effect due to the loss of energy by emission becomes weaker and weaker. These estimates underline that within the framework of this though experiments of radiative equilibrium the amount of the greenhouse effect strongly depends on the planetary emissivity $\varepsilon_E > 0.614$, when the planetary albedo is kept fixed.

To express the uncertainty of $T_e$ due to the uncertainties with which empirical quantities like the planetary albedo and the planetary emissivity are fraught by random errors, we use Gaussian error propagation principles. Thus, the absolute error of temperature $T_e$ is given by

$$\delta T_e = \pm \left\{ \left(\frac{\partial T_e}{\partial \alpha_E}\right)^2 (\delta \alpha_E)^2 + \left(\frac{\partial T_e}{\partial S}\right)^2 (\delta S)^2 + \left(\frac{\partial T_e}{\partial \varepsilon_E}\right)^2 (\delta \varepsilon_E)^2 \right\}^{\frac{1}{2}}, \quad (3.22)$$

where the $\delta \alpha_E$ and $\delta \varepsilon_E$ are the absolute errors (usually the standard deviations) of the planetary albedo and the planetary emissivity, respectively. From this equation we can deduce the relative error given by

$$\frac{\delta T_e}{T_e} = \pm \frac{1}{4} \left\{ \left(\frac{\delta \alpha_E}{1-\alpha_E}\right)^2 + \left(\frac{\delta S}{S}\right)^2 + \left(\frac{\delta \varepsilon_E}{\varepsilon_E}\right)^2 \right\}^{\frac{1}{2}}. \quad (3.23)$$

Figure 9 also illustrates this uncertainty estimated with $\alpha_E = 0.30$, $\delta \alpha_E = 0.02$ (taken from Figure 10), $S = 1367 \text{ W m}^{-2}$, $\delta S = 1.5 \text{ W m}^{-2}$, $\delta \varepsilon_E = 0.025$, and, alternatively, $\delta \varepsilon_E = 0.05$. Thus, in the case of $\varepsilon_E = 0.90$, for instance, the greenhouse effect of $\Delta T \approx 26.2 \text{ K}$ is fraught with an uncertainty of $\pm 2.6 \text{ K}$ for $\delta \varepsilon_E = 0.025$ and $\pm 4.1 \text{ K}$ for $\delta \varepsilon_E = 0.05$, respectively. This temperature difference of $\Delta T \approx 26.2 \text{ K}$ corresponds to a difference in the radiative energy of $\Delta F_{IR} = 111.3 \text{ W m}^{-2}$ with an uncertainty of about $\pm 9.5 \text{ W m}^{-2}$ for $\delta \varepsilon_E = 0.025$ and $\pm 15 \text{ W m}^{-2}$ for $\delta \varepsilon_E = 0.05$, respectively (see Figure 11). This means that the uncertainty in the amount of the so-called greenhouse effect is much larger than the net anthropogenic radiative forcing of $RF = 1.6 \text{ W m}^{-2}$, on average, in 2005 (see Figure 4).

4. **Correct solutions of the conventional global energy balance model of Schneider and Mass for the Earth in the absence of an atmosphere**

*4.1    Analytical and numerical solutions*

Now, it is indispensable to document that the true solution of Eq. (2.1) with Eq. (2.2) leading to

$$R \frac{dT_s}{dt} = (1 - \alpha_E) \frac{S}{4} - \varepsilon_E \sigma T_s^4 \tag{4.1}$$

provides an radiative equilibrium temperature $T_s(\infty)$ which completely agrees with $T_e \cong 254.86 \text{ K}$ as obtained with formula (3.20). Equation (4.1) may also be written as

$$\frac{R}{\varepsilon_E \sigma} \frac{dT_s}{dt} = \frac{(1 - \alpha_E) S}{4 \varepsilon_E \sigma} - T_s^4 = T_e^4 - T_s^4 \quad, \tag{4.2}$$

where the first term of the right-hand-side of this equation is expressed by the radiative equilibrium temperature given formula (3.20). We may do that because, as done before, we again assume that $\alpha_E$, $\varepsilon_E$, and S are independent of time. Separating the variables and integrating from $t = 0$, for which we again have $T_{s0} = T_s(t = 0)$, and t yields

$$\int_{T_{s0}}^{T_s(t)} \frac{dT_s}{T_e^4 - T_s^4} = \frac{\varepsilon_E \sigma}{R} t \quad. \tag{4.3}$$

Since the exact solution of the integral on the right-hand side of this equation reads

$$\int_{T_{s0}}^{T_s(t)} \frac{dT_s}{T_e^4 - T_s^4} = \frac{1}{4 T_e^3} \left\{ \ln \left( \frac{\frac{T_e + T_s(t)}{T_e - T_s(t)}}{\frac{T_e + T_{s0}}{T_e - T_{s0}}} \right) + 2 \arctan \left( \frac{T_e (T_s(t) - T_{s0})}{T_e^2 + T_s(t) T_{s0}} \right) \right\} \quad, \tag{4.4}$$

we finally obtain

$$\frac{1}{4\,T_e^{\,3}} \left\{ \ln\left(\frac{\dfrac{T_e + T_s(t)}{T_e - T_s(t)}}{\dfrac{T_e + T_{s0}}{T_e - T_{s0}}}\right) + 2\arctan\left(\frac{T_e\,(T_s(t) - T_{s0})}{T_e^{\,2} + T_s(t)\,T_{s0}}\right) \right\} = \frac{\varepsilon_E\,\sigma}{R}\,t \quad . \tag{4.5}$$

Unfortunately, this transcendental equation has to be solved for any time $t$ iteratively. Therefore, it is advantageous and convenient to directly solve Eq. (4.1) using a numerical algorithm like Gear's [43] DIFSUB for which we may express Eq. (4.1) in the following manner: $\dot{T}_s = A - B\,T_s^{\,4}$ with $A = (1 - \alpha_E)\,S/(4\,R)$ and $B = \varepsilon_E\,\sigma/R$.

*4.2    Estimates*

Results of the numerical solution of Eq. (4.1) for the aqua planet considering four different values of the water layer thickness $\vartheta$ (25 m, 50 m, 75 m, and 150 m) are illustrated in Figure 12, where, for instance, $T_{s0} = 222\,\text{K}$ and $T_{s0} = 288\,\text{K}$ were chosen as initial values. In all instances the results tend to a value of $T_s(\infty) = 254.86\,\text{K}$, i.e., this value of the numerical solution completely agree with $T_e$. As expected, the fastest response is obtained for $\vartheta = 25\,\text{m}$ with $t \approx 14$ years (considered as the time for which the numerical solution equals $T_e$ with a sufficient degree of accuracy), followed by those of $\vartheta = 50\,\text{m}$ ($t \approx 28$ years), $\vartheta = 75\,\text{m}$ ($t \approx 42$ years), and $\vartheta = 150\,\text{m}$ ($t \approx 84$ years). Note that the values of the e-folding time of response are much smaller, namely $\tau = 0.717$ years for $\vartheta = 25\,\text{m}$, $\tau = 1.43$ years for $\vartheta = 50\,\text{m}$, $\tau = 2.15$ years for $\vartheta = 75\,\text{m}$, and $\tau = 4.30$ years for $\vartheta = 150\,\text{m}$. However, these values are related to the exponential function (2.15) only.

## 5. The formulae of Ångström and Brunt

*5.1 Analytical solutions*

If the down-welling IR radiation is parameterized using either Ångström-type [44, 45] or Brunt-type formulae [45, 46] such a numerical method can be applied to solve Eq. (2.16). These formulae are given by

$$F_{IR\downarrow} = \sigma\, T_L^4 \begin{cases} \left(a_3 - b_3\, 10^{-\gamma e_L}\right) & \text{if Ångström} - \text{type formulae [44, 45]} \\ \left(a_4 + b_4\, e_L^{\frac{1}{2}}\right) & \text{if Brunt} - \text{type formulae [45, 46]} \end{cases} . \qquad (5.1)$$

Here, $T_L$ is the air temperature close to the surface (at the height of 1.5 to 2 m), $e_L$ is the water vapor pressure at the same height either in mm Hg (Ångström) or in hPa (Brunt), and $a_3$, $b_3$, $\gamma$, $a_4$, and $b_4$, are empirical constants listed in Table 2 (see also Kondratyev [34] for additional empirical values). As already shown by Raman [47], Brunt's formula is only a variation of Ångström's formula because

$$F_{IR\downarrow} = \sigma\, T_L^4 \left(a_3 - b_3\, 10^{-\gamma e_L}\right) = \sigma\, T_L^4 \left(\underbrace{a_3 - b_3}_{a_4} + b_3\left(1 - 10^{-\gamma e_L}\right)\right) . \qquad (5.2)$$

In Eq. (5.1) the quantities in the parentheses are notably smaller than unity. This can be explained as follows: The absorption bands of the so-called greenhouse gases have limited ranges of frequencies. The formula for such a limited range also called the filtered spectrum is given by [41, 52, 53]

$$F_F(T) = e_F(X_1, X_2)\sigma\, T^4 \quad , \qquad (5.3)$$

where $e_F(X_1, X_2)$ is defined by

$$e_F(X_1, X_2) = \frac{15}{\pi^4} \int_{X_1}^{X_2} \frac{X^3}{\exp(X) - 1} dX < 1 \qquad (5.4)$$

with $X = h\nu/(kT)$. Here, the subscript F characterized the filtered or finite spectrum. Formula (5.3) describes the fractional emission of a black body due to a finite or filtered spectrum. Note that Eq. (5.4) provides a result even though an emitter/absorber does not exist.

According to Bolz [54], these empirical formulae may be multiplied by a weighting function for both the cloud cover and the cloud type given by

$$f(n) = 1 + K n^{2.5} \quad , \qquad (5.5)$$

where K is a weighting factor that depends on the cloud type. It is ranging from 0.04 for cirrus clouds to 0.24 for stratus clouds; an average value is given by $K \approx 0.22$ [54]. Inserting this formula into Eq. (5.1) yields

$$F_{IR\downarrow} = \sigma T_L^4 \begin{cases} \left(a_3 - b_3 \, 10^{-\gamma e_L}\right)\left(1 + K n^{2.5}\right) & \text{if } [44, 45, 54] \\ \left(a_4 + b_4 \, e_L^{\frac{1}{2}}\right)\left(1 + K n^{2.5}\right) & \text{if } [45, 46, 54] \end{cases} \qquad (5.6)$$

Since the temperatures $T_L$ and $T_s$ can notably differ from each other; we may express the former by the latter according to

$$T_L = T_s - \Delta T \quad , \qquad (5.7)$$

where the deviation $\Delta T$ may be either positive or negative. Thus, the term $\Delta F_{IR} = F_{IR\uparrow}(T_s) - F_{IR\downarrow}$ in Eq. (2.16) results in

$$\Delta F_{IR} = \sigma\, T_s^4 \begin{cases} \left\{\varepsilon_E - \left(1 - 4\,\delta T\right)\left(a_3 - b_3\, 10^{-\gamma e_L}\right)\left(1 + K\, n^{2.5}\right)\right\} & \text{if [44, 45, 54]} \\ \left\{\varepsilon_E - \left(1 - 4\,\delta T\right)\left(a_4 + b_4\, e_L^{\frac{1}{2}}\right)\left(1 + K\, n^{2.5}\right)\right\} & \text{if [45, 46, 54]} \end{cases} \quad , \tag{5.8}$$

where $\delta T = \Delta T/T_s$ is considered as a relative deviation. Since, however, this relative deviation only slightly affects the results it is not further considered, i.e., $T_L$ is approximated by $T_s$. For steady-state conditions (i.e., $dT_s/dt = 0$) Eq. (2.16) provides

$$T_s(\infty) = \left\{\frac{1}{\sigma}\left[\left(1 - \alpha_E - A_a\right)\frac{S}{4} - H - E\right]\right\}^{\frac{1}{4}}$$

$$\times \begin{cases} \left\{\varepsilon_E - \left(a_3 - b_3\, 10^{-\gamma e_L}\right)\left(1 + K\, n^{2.5}\right)\right\}^{-\frac{1}{4}} & \text{if [44, 45, 54]} \\ \left\{\varepsilon_E - \left(a_4 + b_4\, e_L^{\frac{1}{2}}\right)\left(1 + K\, n^{2.5}\right)\right\}^{-\frac{1}{4}} & \text{if [45, 46, 54]} \end{cases} \tag{5.9}$$

*5.2   Estimates*

The surface temperatures provided by formula (5.9) for $\varepsilon_E = 1$, $\alpha_E = 0.30$, $A_a = 0.23$, $H \cong 17\ \text{W m}^{-2}$, $E \cong 80\ \text{W m}^{-2}$ and various values of the water vapor pressure are listed in Table 3. Obviously, these results are ranging from 267.8 K (number II) to 298.7 K.(number VIII) when the effect of water vapor is ignored. With exception of numbers IV, VI, and IX they notably disagree with $\langle T_{ns}\rangle \approx 288\ \text{K}$. For increasing values of the water vapor pressure the surface temperatures become irrelevant because they are much higher than $\langle T_{ns}\rangle$. Including the effect of clouds in the infrared range would still increase these surface temperatures because the function (5.5) becomes larger than unity for $n > 1$. On the other hand, if we ignore the fluxes of sensible and latent heat and the absorption of solar radiation by the atmosphere the surface temperatures are still higher and, of course, senseless.

## 6. **Two-layer energy balance models**

*6.1 Analytical solutions*

Recently, Smith [55] discussed the infrared absorption by the atmosphere to illustrate the so-called greenhouse effect, where he used a two-layer model of radiative equilibrium. Similar models were already discussed, for instance, by Hantel [56] and Kump et al. [57]. In contrast to these models in which the absorption of solar radiation by the atmosphere is not included we consider the more advanced one of Dines [58] (see Figure 13) and Liou [16]. Inserting uniform temperatures for the atmosphere, $T_a$, and the Earth's surface, $T_E$ provides [60]

*Top of the atmosphere:*

$$\left(1 - \alpha_E\right)\frac{S}{4} - \varepsilon_a \sigma T_a^4 - \left(1 - \varepsilon_a\right)\varepsilon_E \sigma T_E^4 = 0 \quad , \tag{6.1}$$

*Earth's surface:*

$$\left(1 - \alpha_E - A_a\right)\frac{S}{4} + \varepsilon_E \varepsilon_a \sigma T_a^4 - \varepsilon_E \sigma T_E^4 = 0 \quad . \tag{6.2}$$

Here, the subscript a characterizes the values for the atmosphere, and, again, the subscript E the values for the Earth's surface. Note that the quantity $1 - \alpha_E - A_a$ is the integral transmissivity; the solar radiation absorbed by the Earth's skin is, therefore, given by $\left(1 - \alpha_E - A_a\right)S/4$. Furthermore, the term $\left(1 - \varepsilon_a\right)\varepsilon_E \sigma T_E^4$ is the terrestrial radiation that is propagating through the atmosphere (it also includes the terrestrial radiation that is passing through the atmospheric window). Moreover, the reflection of infrared radiation at the Earth's surface is included here, but scattering of infrared radiation in the atmosphere is ignored, in accord with Möller [59]. The latter substantially agrees with the fact that in the radiative transfer equation the Planck function is considered as the only source function when a non-scattering medium is in local thermodynamic equilibrium so that a beam of monochromatic intensity passing trough the medium will undergo absorption and emission processes simultaneously, as described by Schwarzschild's equation [16, 61-63]. As before, all properties are considered as uniform, too. The solution of this two-layer model of radiative equilibrium is given by [60]

$$T_a = \left\{ \frac{\left(A_a + \varepsilon_a \left(1 - \alpha_E - A_a\right)\right) S}{4 \, \varepsilon_a \, \sigma \left(1 + \varepsilon_E \left(1 - \varepsilon_a\right)\right)} \right\}^{\frac{1}{4}} \tag{6.3}$$

and

$$T_e = \left\{ \frac{\left(\left(1 + \varepsilon_E\right)\left(1 - \alpha_E\right) - A_a\right) S}{4 \, \varepsilon_E \, \sigma \left(1 + \varepsilon_E \left(1 - \varepsilon_a\right)\right)} \right\}^{\frac{1}{4}} . \tag{6.4}$$

This pair of equation is non-linear with some coupling terms [16]. Choosing $\varepsilon_E = 1$ leads to that of Liou [16]. In the absence of an atmosphere as discussed in sections 3 and 4 the quantities $\varepsilon_E$ and $A_a$ are equal to zero, and Eq. (6.4) will completely agree with Eq. (3.20).

The opposite is true in the case of Budyko's [22] heat balance of the Earth presented in his Figure 1. Budyko even assumed that the solar radiation at the top of the atmosphere is given by $\left(1 - \alpha_E\right) S/4$ with $\alpha_E \cong 0.33$. Then, he argued that the solar radiation reaching the Earth's surface is reduced by atmospheric absorption so that an amount of solar radiation given by $\left(1 - \alpha_E\right)\left(1 - A_a\right) S/4$ is reaching the Earth's surface, where $A_a \cong 0.25$. He further argued that at the Earth's surface a portion of this incident solar radiation is reflected due to an short-wave albedo of the Earth's surface of about $\alpha_S = 0.14$, i.e., he distinguished between the albedo at the top of the atmosphere, $\alpha_E$, that is also affected by the Earth's surface, and the surface albedo $\alpha_S$. Thus, following Budyko [22] the solar radiation absorbed at the Earth's skin would be given by $\left(1-\alpha_E\right)\left(1-A_a\right)\left(1-\alpha_S\right) S/4$. Taking his value of $S \cong 1000$ kcal cm$^{-2}$ yr$^{-1}$ and his values for $\alpha_S = 0.14$, $\alpha_E = 0.33$, and $A_a \cong 0.25$ into account leads to $108$ kcal cm$^{-2}$ yr$^{-1}$ [22]. In the absence of the atmosphere the incident solar radiation absorbed by the Earth's skin would be $\left(1-\alpha_E\right)\left(1-\alpha_S\right) S/4$, where $\alpha_E = \alpha_S$, leading to

$$T_E = \left\{\frac{\left(1-\alpha_S\right)^2 S}{4\,\varepsilon_E\,\sigma}\right\}^{\frac{1}{4}} . \tag{6.5}$$

This formula disagrees with Eq. (3.20), i.e., Budyko's consideration is notably inconsistent.

Figure 3 illustrates that the fluxes of sensible heat, $H$, and latent heat, $E$, are not negligible (see also Figure 1 by Dines [58], here given by Figure 13). Consequently, the radiation flux balance at the Earth's surface has to be replaced by an energy flux balance [60],

$$\left(1-\alpha_E - A_a\right)\frac{S}{4} + \varepsilon_E\,\varepsilon_a\,\sigma\,T_a^{\,4} - \varepsilon_E\,\sigma\,T_E^{\,4} - H - E = 0 \quad, \tag{6.6}$$

to include them, in accord with Trenberth et al. [27]. In doing so, one obtains [60]

$$T_a = \left\{\frac{\left(A_a + \varepsilon_a\left(1-\alpha_E - A_a\right)\right)\frac{S}{4} + \left(1-\varepsilon_a\right)\left(H+E\right)}{\varepsilon_a\,\sigma\left(1+\varepsilon_E\left(1-\varepsilon_a\right)\right)}\right\}^{\frac{1}{4}} \tag{6.7}$$

and

$$T_E = \left\{\frac{\left(\left(1+\varepsilon_E\right)\left(1-\alpha_E\right) - A_a\right)\frac{S}{4} - H - E}{\varepsilon_E\,\sigma\left(1+\varepsilon_E\left(1-\varepsilon_a\right)\right)}\right\}^{\frac{1}{4}} . \tag{6.8}$$

In the absence of the Earth's atmosphere formula Eq. (6.8) will completely agree with Eq. (3.20), too. Note that Arrhenius [64] considered a similar scheme for a column of the atmosphere, i.e., he already included the absorption of solar radiation by atmospheric constituents, and the exchange of heat between the Earth's surface and the atmosphere.

Customarily, the fluxes of sensible and latent heat are expressed by (e.g., [65]):

$$H = -\bar{\rho}\, c_p\, C_h \left(\widehat{u_R} - \widehat{u_s}\right)\left(\widehat{\Theta_R} - \widehat{T_s}\right) = \text{const.} \quad, \tag{6.9}$$

and

$$E = -\bar{\rho}\, L_v\, C_q \left(\widehat{u_R} - \widehat{u_s}\right)\left(\widehat{q_R} - \widehat{q_s}\right) = \text{const.} \quad. \tag{6.10}$$

Here, $C_h$ and $C_q$ are the transfer coefficients for sensible heat and water vapor, respectively. Furthermore, $\widehat{u_R}$ and $\widehat{u_s}$ are the mean values of the wind speed at $z_R$, the outer edge of the atmospheric surface layer (subscript R), and at the Earth's surface (subscript s), where in the case of rigid walls the latter is equal to zero, $\widehat{\Theta_R}$ is the mean potential temperature at $z_R$, $\widehat{T_s}$ is the mean absolute temperature at the surface, and $\widehat{q_r}$ and $\widehat{q_s}$ are the corresponding values of the specific humidity, respectively. Moreover, $\bar{\rho}$ is the mean air density, $c_p$ is the specific heat at constant pressure, and $L_v$ is the mean specific heat of phase transition (e.g., vaporization, sublimation). As expressed by these equations, these fluxes are related to differences of temperature and humidity between a certain reference height, $z_R$, and the Earth's surface. If representative atmospheric values like a uniform temperature and a uniform water vapor pressure are considered, such formulations as given by formulae (6.9) and (6.10) become impracticable. Thus, we have to insert H and E into the energy flux balance equation (6.6) as recommended, for instance, by Trenberth et al. [27].

*6.2    Estimates*

It is obvious that $T_E$ and $T_a$ are dependent on the emissivity values of the Earth and the atmosphere, the absorption and the planetary albedo. Results provided by Eqs. (6.3) and (6.4) using $\alpha_E \cong 0.30$ and some combinations of $\varepsilon_E$ and $\varepsilon_a$, where the absorptivity, $A_a$, is ranging from zero to 0.3, are illustrated in Figure 14. Assuming, for instance, that the atmosphere acts as blackbody emitter leads to an atmospheric temperature of about $T_a \cong 255 \text{ K}$ which is independent of the absorptivity (see Eq. (6.3) for $\varepsilon_a = 1$). Considering, in addition, the Earth as a blackbody emitter provides a surface temperature of about $T_E \cong 303 \text{ K}$ if $A_a$ is assumed to be zero. This value completely agrees with those of Smith [55], Hantel [56], and Kump et al. [57]. Since the global average of temperatures observed in the close vicinity of the Earth's surface corresponds to $\langle T_{ns} \rangle \approx 288 \text{ K}$, this value of $T_E \cong 303 \text{ K}$ confirms that the absorption of solar radiation in the atmosphere causes a decrease of the Earth's surface temperature (the atmospheric temperature increases concurrently). If we additionally assume that also the Earth acts as a blackbody emitter, we will obtain a temperature value for the Earth's surface of $T_E = 288 \text{ K}$ for an absorptivity of $A_a \cong 0.26$. This value is close to those estimated by Budyko [22], Schneider [15], Liou [16], and Trenberth et al. [27] (see also Figure 3). However, as shown in Figure 14 other combinations of $\varepsilon_E$ and $\varepsilon_a$ provide different results. Even though that now $T_a$ grows with an increasing absorptivity, the decrease of $T_E$ is nearly as strong as the increase of $T_a$. Thus, we may conclude that the two-layer model of radiative equilibrium is, in principle, able to provide any pair of results for $T_E$ and $T_a$ we would like.

The results shown in Figure 15 are based on Eqs. (6.7) and (6.8) illustrating the effects owing to the fluxes of sensible and latent heat, where $H = 17 \text{ W m}^{-2}$ and $E = 80 \text{ W m}^{-2}$ are considered, in accord with Trenberth et al. [27]. This figure is based on the same combination of data as used in the case of Figure 14. As illustrated, including these flux values leads to notably lower temperatures at the Earth's surface. Note that in the case of $\varepsilon_E = 0.95$ and $\varepsilon_a = 0.6$ the Earth's surface temperature would be lower than the temperature of the radiative equilibrium of $T_e \approx 255 \text{ K}$ (provided by Eq. (3.20)) for $A_a \geq 0.17$. For this case the Earth's surface temperature would be lower than the temperature of the atmosphere. In such a case it has to be expected that, at least, the sensible heat flux should change its direction. For $\varepsilon_E = 1.0$, $\varepsilon_a = 0.8$, and $A_a = 0.23$ the Earth's surface temperature would only be slightly higher than $T_e$. In this case the temperature of the atmosphere would be $T_a \approx 255 \text{ K}$, i.e., it would correspond to the vertically averaged temperature of the troposphere.

Based on these results we may conclude that this two-layer model of energy flux equilibrium does not support the existence of the so-called atmospheric greenhouse effect as defined, for instance, by the American Meteorological Society (see section 3).

7. **Summary and conclusions**

In this paper we discussed the global energy balance model of Schneider and Mass [2] customarily used to study feedback mechanisms and transient climate response. We pointed out that global heat capacity, $C$, and global inertia coefficient, $R$, differ from each other by a length scale, here the thickness of the water layer of an aqua planet as considered by these authors. Also the feedback parameter, $\lambda$, and the sensitivity parameter, $\lambda^*$, differ from each other. We showed that the latter is the reciprocal of the former, but not identical as stated in the recent NRC report [6]. It is indispensable to pay attention to these issues to avoid the use of inconsistent physical units in the very same equation.

If the effects of absorption of solar radiation by atmospheric constituents and the exchange of sensible and latent heat between the Earth's surface and the atmosphere is ignored the global energy balance model of Schneider and Mass [2] provides results that are ranging from 278.9 K to 299 K for different empirical parameters $a$ and $b$ as suggested by various authors [2, 11, 21-23, 25] (see Table 1), i.e., these temperature values are much higher than the temperature $T_e \approx 255 \text{ K}$ inferred from the planetary radiative equilibrium of the Earth in the absence of an atmosphere (see Eq. (3.20)) even though they notably disagree with $\langle T_{ns} \rangle$. As described in section 3, $T_e$ is customarily used to quantify the so-called greenhouse effect by $\langle T_{ns} \rangle - T_e \approx 33\ \text{K}$. However, the results provided by the Schneider-Mass model become appreciably lower than $T_e$ when $H$ and $E$ are inserted (see Eq. (2.17)). They become still lower if the absorption of solar radiation by atmospheric constituents is included additionally (see Table 1). Since the special case of the planetary radiative equilibrium of the Earth in the absence of an atmosphere is an asymptotic solution of either formula (2.20) or formula (2.21), the global energy balance model of Schneider and Mass [2] does not support the existence of the so-called atmospheric greenhouse effect, as defined, for instance, by the American Meteorological Society (see section 3).

As mentioned before, the planetary radiation balance for an Earth in the absence of an atmosphere is an asymptotic solution of the global energy balance model of Schneider and Mass [2]. This temperature difference is mainly fraught by the inappropriate assumption that the Earth's surface exhibits a uniform surface temperature, $T_e \approx 255\ \text{K}$. Using Gaussian error propagation principles we showed that the inherent uncertainty is much larger than the effect of the net anthropogenic radiative forcing of $RF = 1.6\ \text{W m}^{-2}$, on average, in 2005 that is considered as a perturbation of the natural system.

We further analyzed this perturbation and found that its effect is much smaller than the uncertainty with which the global energy balance climate model of Schneider and Mass [2] is fraught due to the linearization of the power law of Stefan [13] and Boltzmann [14] and the use of an unsuitable reference temperature of $T_r = 0°C$ and the empirical constants $a$ and $\lambda = b$, where the latter is identical with the feedback parameter. Since $RF$ is determined for 2005 and relative to pre-industrial conditions defined at 1750 [9] a portion of this $RF$ is already covered by empirical constants $a$ and $b$ (depending on the period of observation of infrared radiation from which $a$ and $b$ are deduced) completely ignored by the Working Group I of the IPCC [9]. As we argued on the basis of Eq. (2.23), the anthropogenic radiative forcing may produce a decrease of the Earth's surface temperature. Also this result gives evidence that the global energy balance climate model of Schneider and Mass [2] leads to inconsistent physical consequences.

To avoid the linearization of the Stefan-Boltzmann power law we discussed Earth's surface temperature obtained when the down-welling infrared radiation is parameterized using either Ångström-type or Brunt-type formulae and the exchange of sensible and latent heat between the Earth's surface and the atmosphere and the absorption of solar radiation by atmospheric constituents are included. If the water vapor effect is ignored, three of these formulae provide results close to the value of $\langle T_{ns} \rangle \approx 288 \text{ K}$, otherwise the Earth's surface temperatures become too high (see Table 3).

We also discussed the results provided by a Dines-type two-layer energy balance model, eventually adopted by Budyko [22] and others, that provides characteristic uniform temperatures not only for the Earth's surface, but also for the atmosphere. However, for realistic empirical data, also these results do not support the existence of the so-called greenhouse effect. Nevertheless, with exception of the non-existing effect of the infrared radiation reflected back by the atmosphere, Dines-type two-layer energy balance models are still the basis for global annual mean Earth's energy budgets as recently published by Trenberth et al. [27].

Based on our findings we may conclude that it is time to acknowledge that the principles on which global energy balance climate models like that of Schneider and Mass [2] or that of Dines [58] are based have serious physical shortcomings and should not further use to study feedback mechanisms, transient climate response, and the so-called atmospheric greenhouse effect.

**Appendix A: Derivation of equation (2.1)**

To derive the global energy balance model of Schneider and Mass [2] for an aqua planet listed here as Eq. (2.1) we consider the upper layer of such an aqua planet. We assume that this planet can be considered as a sphere with the radius $r_E \cong 6{,}371 \text{ km}$ and the water layer under study as a spherical shell of the thickness $\vartheta$. The volume of this shell is given by

$$V_w = \frac{4}{3} \pi r_E^3 - \frac{4}{3} \pi (r_E - \vartheta)^3 = \frac{4}{3} \pi r_E^3 \left\{ 1 - \left(1 - \frac{\vartheta}{r_E}\right)^3 \right\} \quad . \tag{A.1}$$

Since $\vartheta/r_E \ll 1$ the term $(1 - \vartheta/r_E)^3$ can be approximated by $1 - 3\,\vartheta/r_E$, and, hence, the volume by

$$V_w \cong 4 \pi r_E^2 \, \vartheta \quad . \tag{A.2}$$

For the purpose of simplification, we assume that $V_w$ is independent of time. The surface of this water shell is given by $A_w = A_O + A_I = 4 \pi r_E^2 + 4 \pi (r_E - \vartheta)^2 = 4 \pi r_E^2 \left\{ 1 + (1 - \vartheta/r_E)^2 \right\}$, where $A_O$ is the outer surface that is congruent with the surface of the aqua planet, and $A_I$ is the inner surface.

The temporal change of the energy, $E$, in this volume can be expressed by

$$\frac{dE}{dt} = \frac{d}{dt} \int_{V_w} \rho_w \, e \, dV \quad . \tag{A.3}$$

Here, $t$ is time, $\rho_w$ is the density of water, and $e$ is the specific internal energy. The latter may be expressed by $e = c_w \, T_w$, where $c_w$ is the specific heat of water, and $T_w$ is the water temperature. Multiplying $\rho_w$ and $c_w$ with each other provides the heat capacity $C = \rho_w \, c_w$.

The integral term of the right-hand side of Eq. (A.3) is given by

$$\frac{d}{dt} \int_{V_w} C \, T_w \, dV = - \int_{A_w} \mathbf{F}_{e,i} \cdot \mathbf{n} \, dA + \int_{V_w} \sigma_e \, dV \quad , \tag{A.4}$$

where the $\mathbf{F}_{e,i}$ are the various energy fluxes, $\mathbf{n}$ is the unit normal (counted positive from inside to outside of the volume. The surface integral describes the exchange of energy between the volume of the water shell and its surroundings. The term $\sigma_e$ characterizes a possible gain or loss of energy inside of the volume, for instance, owing to friction (only gain) and chemical processes. Equation (A.4) is called the integral energy balance equation.

If we assume that $\sigma_e$ is negligible, Eq. (A.4) can be simplified. In doing so, we obtain

$$\frac{d}{dt} \int_{V_w} C\, T_w\, dV = - \int_{A_w} \mathbf{F}_{e,i} \cdot \mathbf{n}\, dA \quad . \tag{A.5}$$

Since $V_w$ is assumed to be independent of time, the left-hand side of this equation can be expressed by

$$\frac{d}{dt} \int_{V_w} C\, T_w\, dV = V_w \frac{d}{dt} \frac{1}{V_w} \int_{V_w} C\, T_w\, dV = V_w \frac{d}{dt} \langle C\, T_w \rangle_V \quad , \tag{A.6}$$

where

$$\langle \Psi \rangle_V = \frac{1}{V_w} \int_{V_w} \Psi\, dV \tag{A.7}$$

defines the volume average. Here, $\Psi$ is an arbitrary quantity.

Since we have only to consider the energy fluxes across the outer and the inner surface of the water shell, the surface integral in Eq. (A.5) can be written as

$$\int_{A_w} \mathbf{F}_{e,i} \cdot \mathbf{n}\, dA = \int_{A_O} \mathbf{F}_{e,i} \cdot \mathbf{n}\, dA + \int_{A_I} \mathbf{F}_{e,i} \cdot \mathbf{n}\, dA \quad . \tag{A.8}$$

In accord with Schneider and Mass [2] we first assume that (a) only the incoming solar radiation penetrating the water shell, the down-welling infrared radiation and emitted infrared radiation at the outer surface of the water layer have to be considered, i.e., any exchange of sensible and latent heat between the water layer and the atmosphere is ignored, (b) absorption of solar radiation by atmospheric constituents plays no role, and (c) no energy exchange between the water shell under study and the underlying water or the ocean floor does exist. Thus, Eq. (A.8) may be written as

$$\int_{A_w} \mathbf{F}_{e,i} \cdot \mathbf{n} \, dA = \int_{A_O} \mathbf{F}_{e,i} \cdot \mathbf{n} \, dA$$
$$= -r_E^2 \int_\Omega \left\{ \left(1 - \alpha(\Theta_0, \theta, \phi)\right) F \cos \Theta_0 + F_{IR\downarrow}(\theta, \phi) - F_{IR\uparrow}(T_s(\theta, \phi)) \right\} d\Omega \quad . \quad (A.9)$$

Here, $F = (d_0/d)^2 S$ is the solar irradiance reaching the top of the atmosphere approximated by the solar constant, $S$ (see Eq. (3.5)), $\alpha(\Theta_0, \theta, \phi)$ is the albedo, $F_{IR\downarrow}(\theta, \phi)$ is the down-welling infrared radiation, and $F_{IR\uparrow}(T_s(\theta, \phi))$ is the emitted infrared radiation that depends on the temperature of the outer surface of the water shell, $T_s(\theta, \phi)$. Furthermore, $\Omega = 4\pi$ is the solid angle of a sphere, $d\Omega = \sin\theta \, d\theta \, d\phi$ is the differential solid angle, $\theta$ and $\phi$ are the zenith and azimuthal angles, respectively (see Figure 6). Moreover, the local zenith angle of the Sun's center, $\Theta_0$, can be determined using $\cos \Theta_0 = \cos\theta \sin\delta + \sin\theta \cos\delta \cos h$, where $\delta$ is the solar declination angle, and $h$ is the hour angle from the local meridian (see also subsection 3.1). Using the surface average of the globe defined by

$$\langle \Phi \rangle_A = \frac{r_E^2 \int_\Omega \Phi \, d\Omega}{r_E^2 \int_\Omega d\Omega} = \frac{1}{4\pi} \int_\Omega \Phi \, d\Omega \quad , \quad (A.10)$$

where $\Phi$ is an arbitrary quantity, yields

$$\int_{A_w} \mathbf{F}_{e,i} \cdot \mathbf{n} \, dA = -\left\{ \left\langle \left(1 - \alpha(\Theta_0, \theta, \phi)\right) S \cos \Theta_0 \right\rangle_A + \left\langle F_{IR\downarrow}(\theta, \phi) \right\rangle_A - \left\langle F_{IR\uparrow}(T_s(\theta, \phi)) \right\rangle_A \right\} A_O \quad . \quad (A.11)$$

Thus, combining Eqs. (A.5), (A.6), and (A.11) provides

$$\vartheta \frac{d}{dt} \langle C T_w \rangle_V = \left\langle \left(1 - \alpha(\Theta_0, \theta, \phi)\right) S \cos \Theta_0 \right\rangle_A + \left\langle F_{IR\downarrow}(\theta, \phi) \right\rangle_A - \left\langle F_{IR\uparrow}(T_s(\theta, \phi)) \right\rangle_A \quad , \quad (A.12)$$

where $\vartheta = V_w/A_O$.

If we assume that the water is well-mixed so that its surface temperature $T_s$ equals the temperature $T_w$ for the entire water shell, i.e., $T_s(\theta, \phi) = T_s = T_w$, and $C$ is constant in space and time the right-hand side of Eq. (A.12) can be written as

$$\vartheta \frac{d}{dt}\langle C\, T_w \rangle_V = C\, \vartheta \frac{dT_s}{dt} \quad . \tag{A.13}$$

The global average of the incoming solar radiation penetrating the water shell is given by (see also Eqs. (3.9) and (3.14))

$$\langle (1 - \alpha(\Theta_0, \theta, \phi))\, S \cos \Theta_0 \rangle_A = (1 - \alpha_E)\frac{S}{4} \quad , \tag{A.14}$$

where a constant albedo for the entire planet, i.e., $\alpha(\Theta_0, \theta, \phi) = \alpha_E$ $\alpha_E$ is assumed, and either $\delta = \pi/2$ (leading to Eq. (3.9)) or $\delta = 0$ (leading to Eq. (3.14)) is chosen. To get an expression for the global average of the emitted infrared radiation we again assume that the surface temperature is homogeneously distributed, i.e., $T_s(\theta, \phi) = T_s$. In doing so, we obtain (see also subsection 3.2)

$$\langle F_{IR\uparrow}(T_s) \rangle_A = \varepsilon_E\, \sigma\, T_s^4 \quad , \tag{A.15}$$

where $\varepsilon(\theta, \phi) = \varepsilon_E$ denotes the planetary emissivity also considered as independent of the location. The down-welling infrared radiation may be dealt with in a similar manner.

Since all flux quantities are considered as globally averaged values, we may be dropping the average sign, $\langle \ldots \rangle_A$. In doing so, we obtain.

$$R\frac{dT_s}{dt} = (1 - \alpha_E)\frac{S}{4} - \Delta F_{IR} \quad , \tag{A.16}$$

where $\Delta F_{IR}$ is given by $\Delta F_{IR} = F_{IR\uparrow}(T_s) - F_{IR\downarrow}$, and the planetary thermal inertia coefficient is defined by $R = C\vartheta$. Equation (2.1) is identical with Eq. (A.16). For $\rho_w = 1000 \text{ kg m}^{-3}$, $c_w = 4184 \text{ J kg}^{-1} \text{ K}^{-1}$, and $\vartheta = 75 \text{ m}$ we obtain $R = 3.138 \cdot 10^8 \text{ J m}^{-2} \text{ K}^{-1}$ which completely agrees with $R = 7500 \text{ cal cm}^{-2} \text{ K}^{-1}$ used by Schneider and Mass [2].

As mentioned before, no exchange of sensible and latent heat between the water layer of the aqua planet and the atmosphere is included in the global energy balance model of Schneider and Mass[2]. Furthermore, the absorption of solar radiation by atmospheric constituents is ignored. If the atmospheric absorption, $A_a S$, and the vertical components of the fluxes of sensible heat, $H$, and latent heat, $E = L_v(T_s)Q$, at the outer surface of the water shell are inserted into Eq. (A.16), where $L_v(T_s)$ is the latent heat of vaporization, often considered as dependent on the surface temperature, $T_s$, and $Q$ is the water vapor flux, we will obtain (see Figure 16)

$$R \frac{dT_s}{dt} = \left(1 - \alpha_E - A_a\right) \frac{S}{4} - H - E - \Delta F_{IR} \quad . \tag{A.17}$$

Again, all flux quantities are considered as globally averaged values.

**References**

[1] Manabe S, Stouffer R J 2007 Role of ocean in global warming *J. Meteor. Soc. Japan* **85B**, 385-403.

[2] Schneider S H, Mass C 1975 Volcanic dust, sunspots, and temperature trends *Science* **190**, 741-746.

[3] Schwartz S E 2007 Heat capacity, time constant, and sensitivity of Earth's climate system *J. Geophys. Res.* **112**, D24S05 doi:10.1029/2007JD008746.

[4] Hansen J, Lacis A, Rind D, Russell G, Stone P, Fung I, Ruedy R, Lerner J 1984 Climate sensitivity: Analysis of feedback mechanisms eds. Hansen J E, Takahashi T, *Climate Processes and Climate Sensitivity,* Maurice Ewing Ser. No. 5, (Washington, D.C., American Geophysical Union) pp. 130-163.

[5] Dickinson R E 1985 Climate sensitivity. In: Manabe S *Issues in Atmospheric and Oceanic Modeling*, Advances in Geophysics, Vol. 28 (Orlando, FL, Academic Press) pp. 99-129.

[6] National Research Council (NRC) 2005 *Radiative Forcing of Climate Change: Expanding the Concept and Addressing Uncertainties* (Natl. Acad., Washington, D.C.).

[7] Pielke sr. R A, Davey C A, Niyogi D, Fall S, Steinweg-Woods J, Hubbard K, Lin X, Cai M, Lim Y-K, Li H, Nielsen-Gammon J, Gallo K, Hale R, Mahmood R, Foster S, McNider R T, Blanken P 2007 Unresolved issues with the assessment of multidecadal global land surface temperature trends *J. Geophys. Res.* **112**, D24S08, doi:10.1029/ 2006JD008229.

[8] Schneider S H, Dickinson R E 1974 Climate modeling *Rev. Geophys. Space Phys.* **12**, 447-493.

[9] Forster P et al. 2007 Changes in Atmospheric Constituents and in Radiative Forcing eds. Solomon S, Qin D, Manning M, Chen Z, Marquis M, Averyt K B, Tignor M, Miller H L *Climate Change 2007: The Physical Science Basis - Contribution of Working Group I to*

*the Fourth Assessment Report of the Intergovernmental Panel on Climate Change* (Cambridge/New York, Cambridge University Press) pp. 129-234.

[10]     Ramaswamy V et al. 2001 Radiative forcing of climate change eds. Houghton J T et al. *Climate Change 2001: The Scientific Basis - Contribution of Working Group I to the Third Assessment Report of the Intergovernmental Panel on Climate Change* (Cambridge/New York, Cambridge University Press) pp. 349-416.

[11]     Kiehl J T 1992 Atmospheric general circulation modeling ed. Trenberth K E *Climate System Modeling* (Cambridge/New York, Cambridge University Press) pp. 319-369.

[12]     Ramanathan V, Callis L, Cess R, Hansen J, Isaksen I, Kuhn W, Lacis A, Luther F, Mahlman J, Reck R, Schlesinger M 1987 Climate-chemical interactions and effects of changing atmospheric trace gases *Reviews of Geophysics* **25** (7), 1441-1482.

[13]     Stefan J 1879 Über die Beziehung zwischen der Wärmestrahlung und der Temperatur *Wiener Ber. II* **79**, 391-428 (in German).

[14]     Boltzmann L 1884 Ableitung des Stefan'schen Gesetzes, betreffend die Abhängigkeit der Wärmestrahlung von der Temperatur aus der electromagnetischen Lichttheorie *Wiedemann's Annalen* **22**, 291-294 (in German).

[15]     Schneider S H 1992 Introduction to climate modeling ed. Trenberth K E *Climate System Modeling* (Cambridge/New York, Cambridge University Press) pp. 3-26.

[16]     Liou K N 2002 *An Introduction to Atmospheric Radiation - Second Edition* (San Diego, CA, Academic Press).

[17]     Petty G W 2004 *A First Course in Atmospheric Radiation* (Madison, WI, Sundog Publishing).

[18]     Bohren C F, Clothiaux E E 2006 *Fundamentals of Atmospheric Radiation* (Berlin, Germany, Wiley-VCH).

[19]     Aldrich R B, Hoover W H 1952 The solar constant *Science* **116**, 3.

[20]     Laue E G, Drummond A J 1968 Solar constant: First direct measurements *Science* **116**, 888 – 891.

[21]   Budyko M I 1969 The effect of solar radiation variations on the climate of the Earth *Tellus* **21**, 611-619.

[22]   Budyko M I 1977 *Climatic Change* (Washington, D.C., American Geophysical Union).

[23]   Manabe S, Wetherald R T 1967 Thermal equilibrium of the atmosphere with a given distribution of relative humidity *J. Atmos. Sci.* **24**, 241-259.

[24]   Sellers W D 1969 A global climatic model based on the energy balance of the earth-atmosphere system *J. Appl. Meteor.* **8**, 392-400.

[25]   North G R 1975 Theory of energy-balance climate models *J. Atmos. Sci.* **32**, 2033-2045.

[26]   Lange H-J 2007 *Wetter und Klima im Phasenraum* (Summary of two presentations to "climate in the phase space" http://hajolange.de/).

[27]   Trenberth K E, Fasullo J T, Kiehl J 2009 Earth's global energy budget *Bull. Amer. Meteor. Soc.* 311-323.

[28]   Kramm G, Herbert F 2009 Similarity hypotheses for the atmospheric surface layer expressed by non-dimensional characteristic invariants - A review *The Open Atmospheric Science Journal* **3**, 48-79.

[29]   Kasten F, Raschke E 1974 Reflection and transmission terminology by analogy with scattering *Appl. Optics* **13**, 450-464.

[30]   Haltiner G J, Martin F L 1957 *Dynamical and Physical Meteorology* (New York/Toronto/London, McGraw-Hill).

[31]   Möller F 1974 *Einführung in die Meteorologie*, Bd.2 (Mannheim/Wien/Zürich, Bibliographisches Institut, in German).

[32]   Iqbal M 1983 *An Introduction to Solar Radiation* (Academic Press Canada).

[33]   Vardavas I M, Taylor F W 2007 *Radiation and Climate* (Oxford, U.K., Oxford University Press).

[34]   Kondratyev K YA 1969 *Radiation in the Atmosphere* (New York/London, Academic Press).

[35] Kramm G, Dlugi R, Zelger M 2008 On the recognition of fundamental physical principles in recent atmospheric-environmental studies *J. Calcutta Math. Soc.* **4** (1 & 2), 31-55.

[36] Hartmann D L 1994 *Global Physical Climatology* (San Diego, CA, Academic Press).

[37] Peixoto J P, Oort A H 1992 *Physics of Climate* (New York, Springer).

[38] Cremers C J, Birkebak R C, White J E 1971 Lunar surface temperature at tranquility base *AIAA Journal* **9**, 1899-1903.

[39] Mukai T, Tanaka M, Ishimoto H, Nakamura R 1997 Temperature variations across craters in the polar regions of the Moon and Mercury *Adv. Space Res.* **19**, 1497–1506.

[40] Vasavada A R, Paige D A, Wood S E 1999 Near-surface temperatures on Mercury and the Moon and the stability of polar ice deposits *Icarus,* **141**, 179–193.

[41] Gerlich G, Tscheuschner R D 2009 Falsification of the atmospheric $CO_2$ greenhouse effects within the frame of physics *Int. J. Mod. Phys.* **B23**, 275-364.

[42] Möller F 1964 Optics of the lower atmosphere *Applied Optics* **3** (2), 157-166.

[43] Gear C W *Numerical Initial Value Problems in Ordinary Differential Equations* (Englewood Cliffs, NJ, Prentice-Hall Inc.).

[44] Ångström A 1916 Über die Gegenstrahlung der Atmosphäre *Met. Zeitschrift* **33**, 529-538 (in German).

[45] Bolz H M, Falckenberg G 1949 Neubestimmung der Konstanten der Ångströmschen Strahlungsformel *Z. Meteorologie* **3**, 97-100 (in German).

[46] Brunt D 1932 Notes on radiation in the atmosphere *Q. J. R. Meteorol. Soc.* **58**, 389–420.

[47] Raman P K 1935 Heat radiation from the clear atmosphere at night *Proc. Indian Acad. Sci.* **1**, 815-821.

[48] Ångström A 1929 Über Variationen der atmosphärischen Temperaturstrahlung und ihren Zusammenhang mit der Zusammensetzung der Atmosphäre *Gerlands Beitr. Geophys.* **21**, 145-161 (in German).

[49] Ångström A 1933. Linke's Meteorologisches Taschenbuch (Leipzig, Germany, Akademische Verlagsgesellschaft, in German, as cited by [45]).

[50] Dines W H, Dines L H G 1927 Monthly mean values of radiation from various parts of the sky at Benson, Oxfordshire *Mem. Roy. Met. Soc.* **2**, No. 11.

[51] Philipps H 1940 Zur Theorie der Wärmestrahlung in Bodennähe *Gerlands Beitr. Geophys.* **56**, 229-319 (in German).

[52] Modest M F 2003 *Radiative Heat Transfer* (Amsterdam/Boston/London/New York/ Oxford/Paris/San Diego/San Francisco/Singapore/Sydney/Tokyo, Academic Press).

[53] Kramm G, Mölders N 2009 Planck's blackbody radiation law: Presentation in different domains and determination of the related dimensional constants *J. Calcutta Math. Soc.* **5** (1 & 2), 37-61.

[54] Bolz H M 1949 Die Abhaengigkeit der infraroten Gegenstrahlung von der Bewölkung *Z. Meteorologie* 3, 201-203 (in German).

[55] Smith A P 2008 Proof of the atmospheric greenhouse effect (http://arxiv.org/abs/0802.4324).

[56] Hantel M 1997 Klimatologie *Bergmann, Schaefer – Lehrbuch der Experimentalphysik, Band 7, Erde und Planeten* (Berlin/New York, Walter de Gruyter) 311-426 pp.

[57] Kump L R, Kasting J F, Crane R G 2004 *The Earth System* (Upper Saddle River, NJ, Pearson Education).

[58] Dines W H 1917 The heat balance of the atmosphere *Quart. J. Roy. Met. Soc.* **43**, 151-158.

[59] Möller F 1973 Geschichte der meteorologischen Strahlungsforschung *promet* **2** (in German).

[60] Kramm G, Dlugi R, Zelger M 2009 Comments on the "Proof of the atmospheric greenhouse effect" by Arthur P. Smith (http://arxiv.org/abs/0904.2767v3).

[61] Chandrasekhar S 1960 *Radiative Transfer* (New York, Dover Publications).

[62] Goody R M, Yung Y L 1989 *Atmospheric Radiation* (New York/Oxford, Oxford University Press).

[63] Lenoble J 1993 *Atmospheric Radiative Transfer* (Hampton, VA, A. Deepak Publishing).

[64] Arrhenius S 1896 On the influence of carbonic acid in the air upon the temperature of the ground *Philosophical Magazine* **41**, 237-275.

[65] Pal Arya S 1988 *Introduction to Micrometeorology* (San Diego, Academic Press).

Table 1: Earth's surface temperatures provided by Eqs. (2.6) and (2.17). In the case of Eq. (2.17) the calculations were performed without and with the absorption of solar radiation by atmospheric constituents.

| a $(W\ m^{-2})$ | b $(W\ m^{-2}\ K^{-1})$ | Eq (2.6) $T_s(\infty)$ (K) | Eq. (2.17)[1] $T_s(\infty)$ (K) | Eq. (2.17) $T_s(\infty)$ (K) | $\Delta T_s$ (K) | Author(s) |
|---|---|---|---|---|---|---|
| 226.0 | 2.26 | 279.0 | 236.1 | 201.3 | 0.71 | [21-23] |
| 201.5 | 1.45 | 299.2 | 232.3 | 178.1 | 1.1 | [2] |
| 211.2 | 1.55 | 291.2 | 228.7 | 177.9 | 1.0 | [11, 25] |

[1]) No absorption of solar radiation by atmospheric constituents ($A_a = 0$).

Table 2: Constants of the Ångström-type and Brunt-type formulae (adopted from Bolz and Falckenberg [47]).

| Number | $a_3$ | $b_3$ | $\gamma$ | $a_4$ | $b_4$ | Year | Author | Remarks |
|---|---|---|---|---|---|---|---|---|
| I | 0.79 | 0.26 | 0.069 | - | - | 1916 | [44] | Nighttime values |
| II | 0.75 | 0.32 | 0.069 | - | - | 1929 | [48] | Nighttime values |
| III | - | - | - | 0.47 | 0.072 | 1932 | [45] | Nighttime values |
| IV | 0.806 | 0.236 | 0.069 | - | - | 1933 | [49] | Nighttime values |
| V | 0.77 | 0.28 | 0.075 | - | - | 1935 | [47] | Nighttime values |
| VI | - | - | - | 0.57 | 0.034 | 1935 | [47] | Nighttime values |
| VII | - | - | - | 0.52 | 0.065 | | [50] | Nighttime values |
| VIII | 0.78 | 0.148 | 0.068 | - | - | 1940 | [51] | Nighttime values |
| IX | 0.820 | 0.250 | 0.126 | - | - | 1948 | [45] | Daytime and nighttime values |

Table 3: Earth's surface temperatures provided by Ångström-type and Brunt-type formulae for different water vapor pressures.

| Number | $e_L$ (hPa) | | | | |
|---|---|---|---|---|---|
| | 0 | 5 | 10 | 15 | 20 |
| I | 281.0 | 301.8 | 317.3 | 327.9 | 334.5 |
| II | 267.8 | 287.9 | 303.1 | 313.5 | 320.0 |
| III | 272.7 | 298.5 | 313.8 | 328.7 | 344.5 |
| IV | 287.3 | 308.4 | 324.1 | 334.7 | 341.4 |
| V | 275.3 | 297.0 | 312.7 | 322.7 | 328.7 |
| VI | 287.3 | 301.6 | 308.7 | 314.8 | 320.4 |
| VII | 279.5 | 305.9 | 321.4 | 336.6 | 352.7 |
| VIII | 298.7 | 313.8 | 324.1 | 330.6 | 334.5 |
| IX | 287.3 | 324.5 | 344.4 | 352.6 | 355.6 |

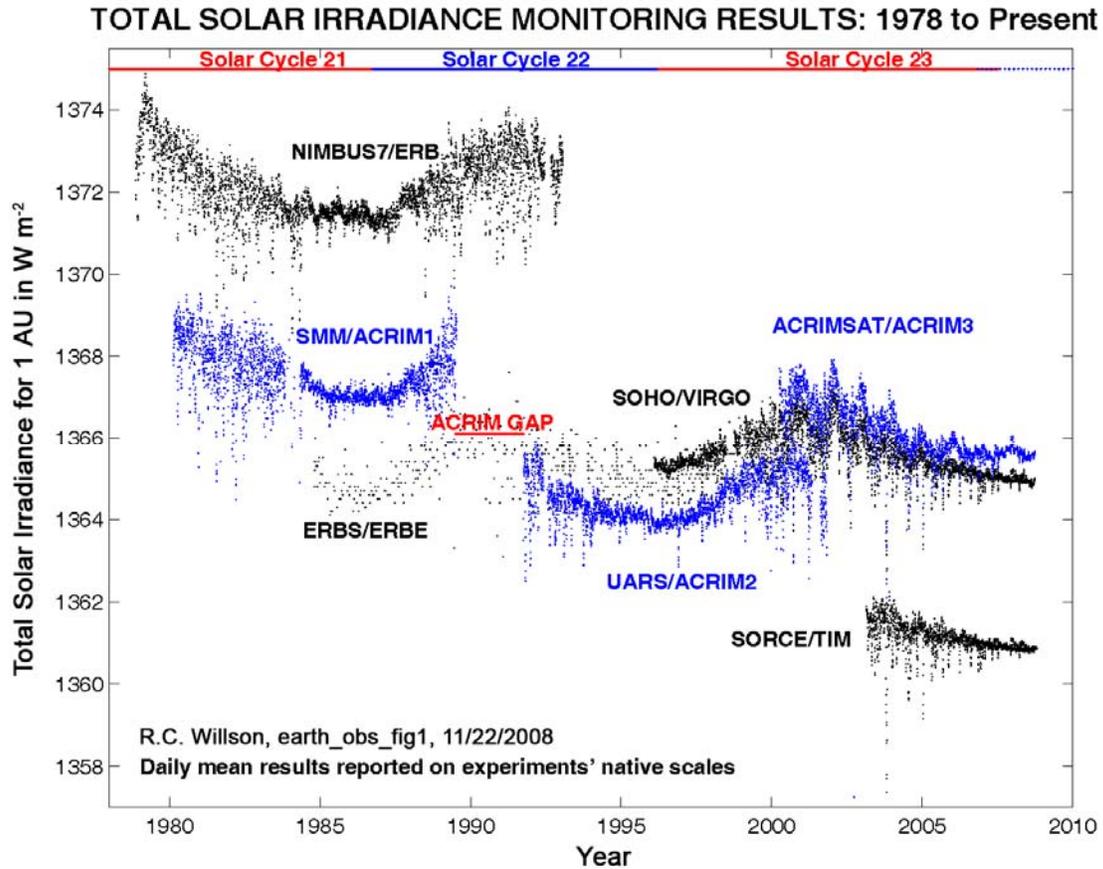

**Figure 1:** Satellite observations of total solar irradiance. It comprises of the observations of seven independent experiments: (a) Nimbus7/Earth Radiation Budget experiment (1978 - 1993), (b) Solar Maximum Mission/Active Cavity Radiometer Irradiance Monitor 1 (1980 - 1989), (c) Earth Radiation Budget Satellite/Earth Radiation Budget Experiment (1984 - 1999), (d) Upper Atmosphere Research Satellite/Active cavity Radiometer Irradiance Monitor 2 (1991 - 2001), (e) Solar and Heliospheric Observer/Variability of solar Irradiance and Gravity Oscillations (launched in 1996), (f) ACRIM Satellite/Active cavity Radiometer Irradiance Monitor 3 (launched in 2000), and (g) Solar Radiation and Climate Experiment/Total Irradiance Monitor (launched in 2003). The figure is based on Dr. Richard C. Willson's earth_obs_fig1, updated on November 22, 2008 (see http://www.acrim.com/).

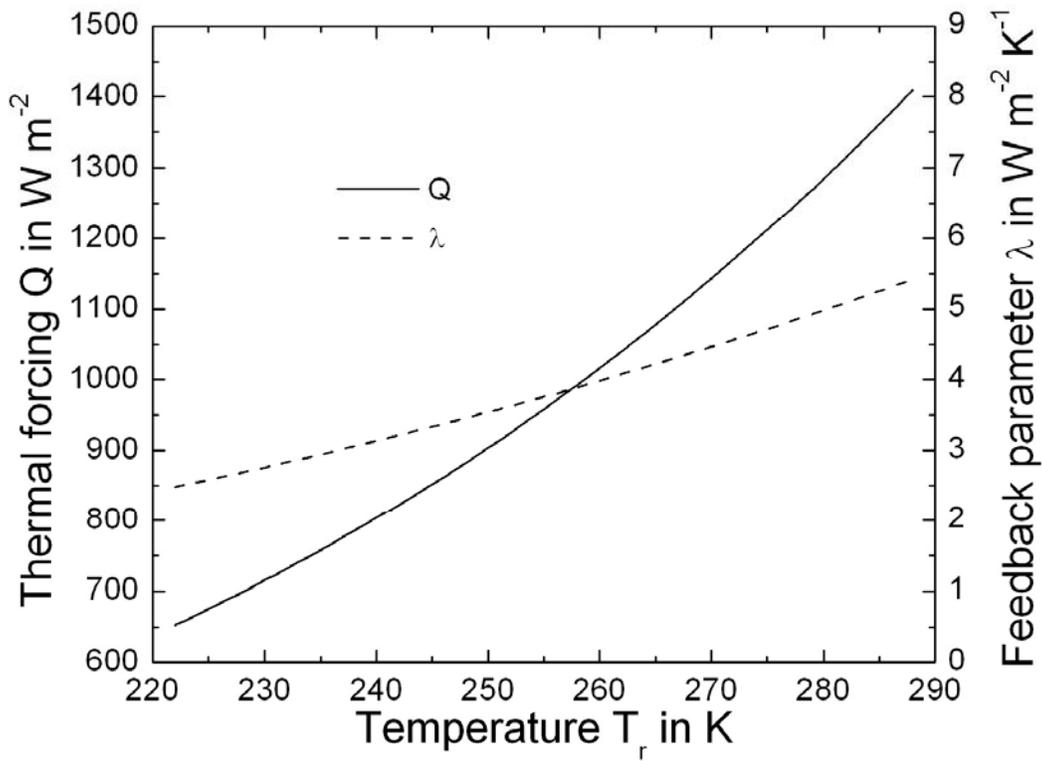

**Figure 2:** Thermal forcing Q and feedback parameter λ versus reference temperature $T_r$ used by the linearization of the power law of Stefan [13] and Boltzmann [14] (see formulae (2.10) to (2.12)).

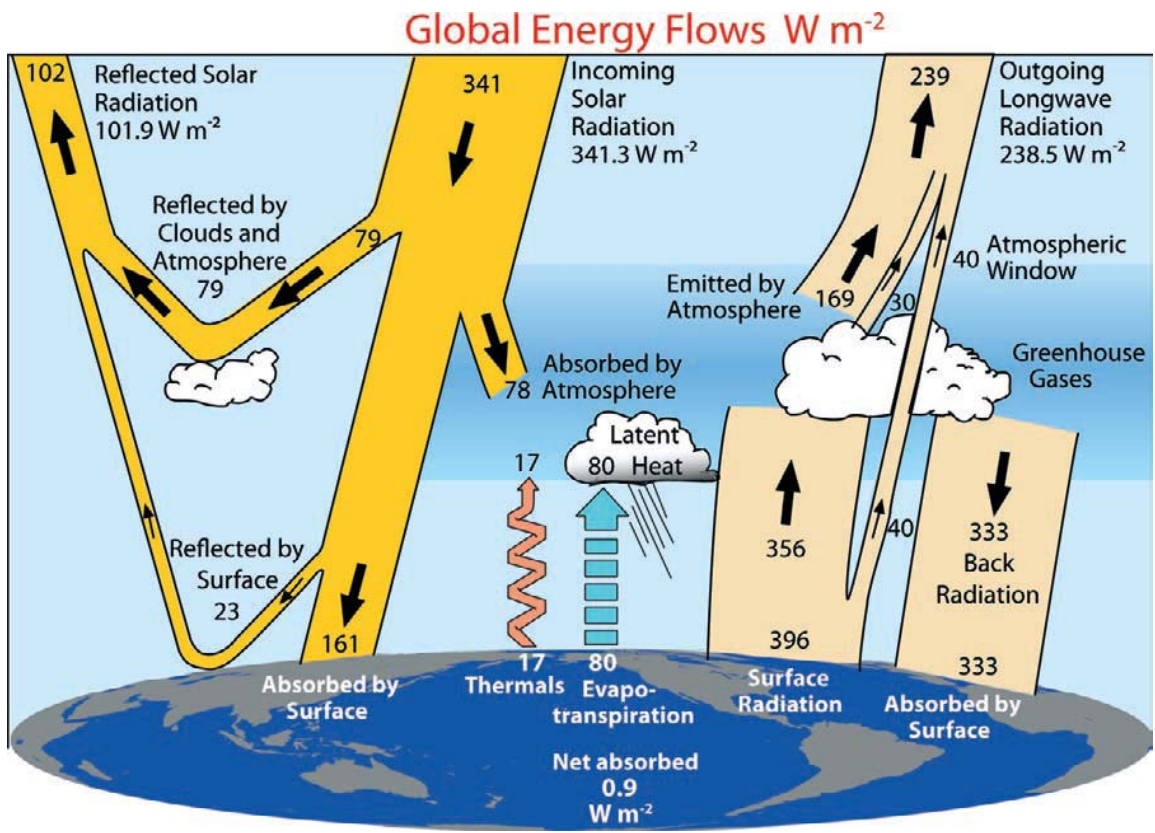

**Figure 3**: The global annual mean Earth's energy budget for the Mar 2000 to May 2004 period (W m$^{-2}$). The broad arrows indicate the schematic flow of energy in proportion to their importance (adopted from Trenberth et al. [27]).

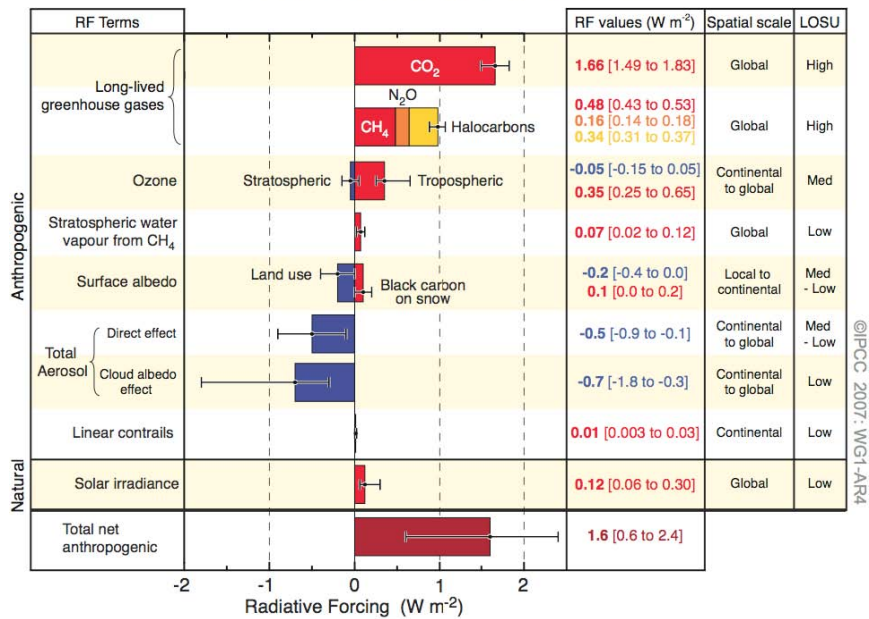

**Figure 4:** Global-average radiative forcing (RF) estimates and ranges in 2005 for anthropogenic carbon dioxide ($CO_2$), methane ($CH_4$), nitrous oxide ($N_2O$) and other important agents and mechanisms, together with the typical geographical extent (spatial scale) of the forcing and the assessed level of scientific understanding (LOSU). The net anthropogenic radiative forcing and its range are also shown. These require summing asymmetric uncertainty estimates from the component terms, and cannot be obtained by simple addition. Additional forcing factors not included here are considered to have a very low LOSU. Volcanic aerosols contribute an additional natural forcing but are not included in this figure due to their episodic nature. Range for linear contrails does not include other possible effects of aviation on cloudiness (adopted from Forster et al. [9]).

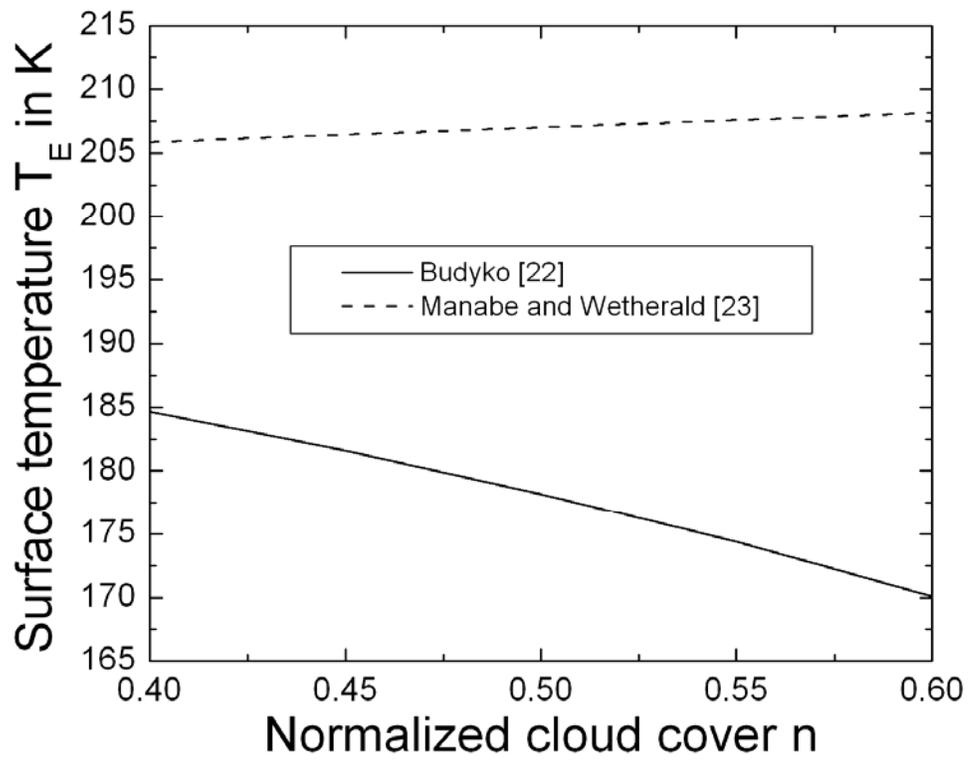

**Figure 5:** Surface temperature vs. normalized cloud cover for the empirical formulae of Budyko [22] and Manabe and Wetherald [23].

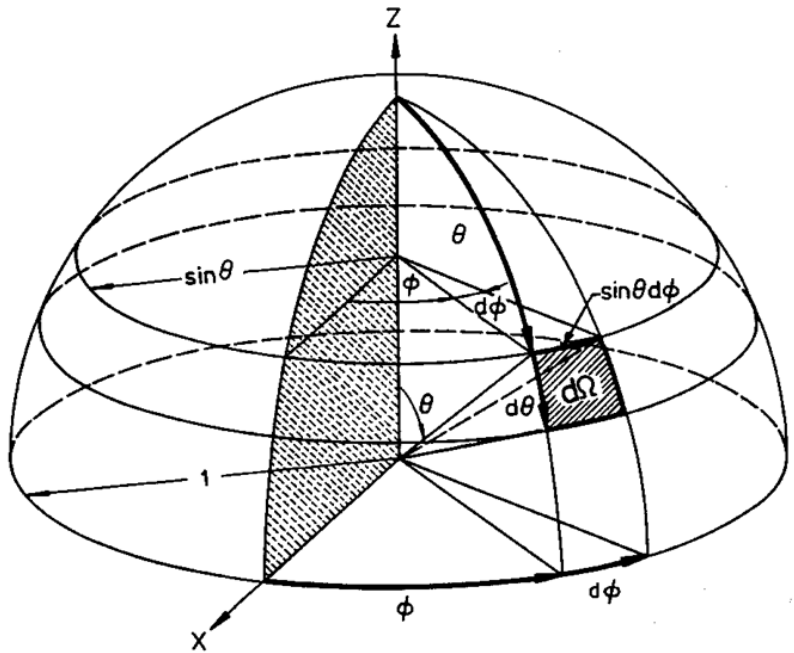

**Figure 6:** Mathematical representation of the solid angle. Here, $d\Omega = \sin\theta\, d\theta\, d\phi$ is the differential solid angle, where $\theta$ and $\phi$ are the zenith and azimuthal angles, respectively (adopted from Kasten and Raschke [29]).

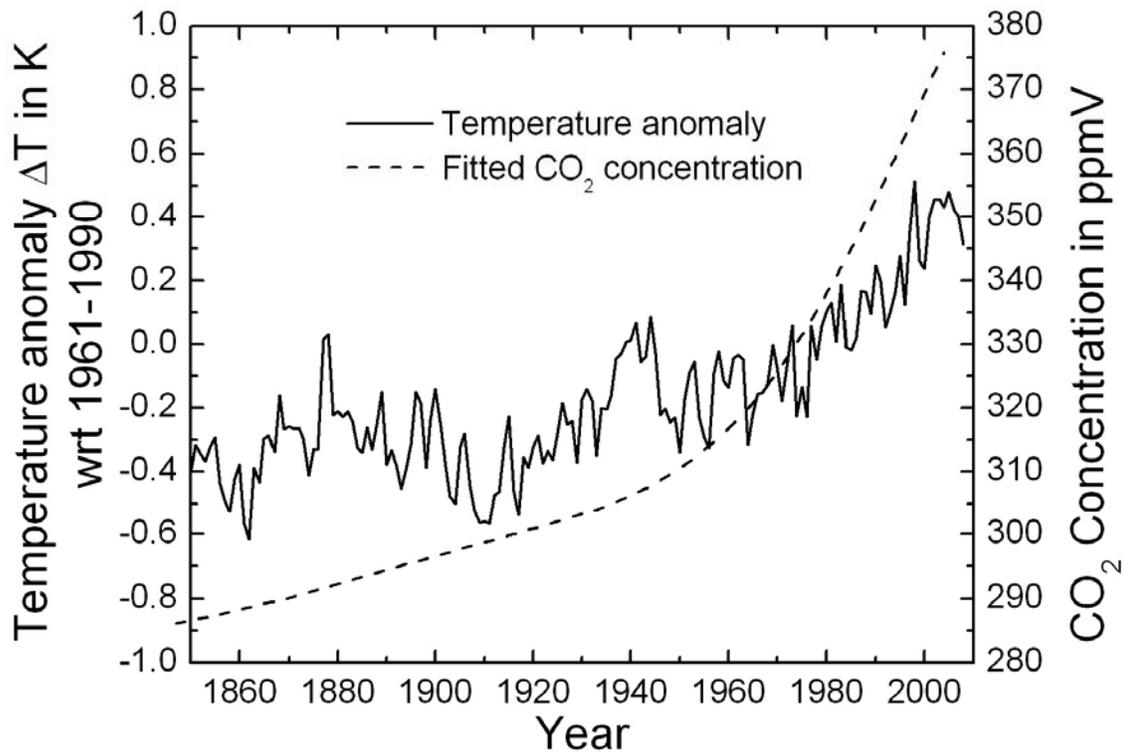

**Figure 7**: The increase of the globally averaged near-surface temperature (annual mean) expressed by the temperature anomaly with respect to the period 1961-1990. These temperature anomaly data were adopted from the Hadley Centre for Climate Prediction and Research, MetOffice, UK. The fit of the atmospheric carbon dioxide ($CO_2$) concentration adopted from Kramm et al. [35] is also shown.

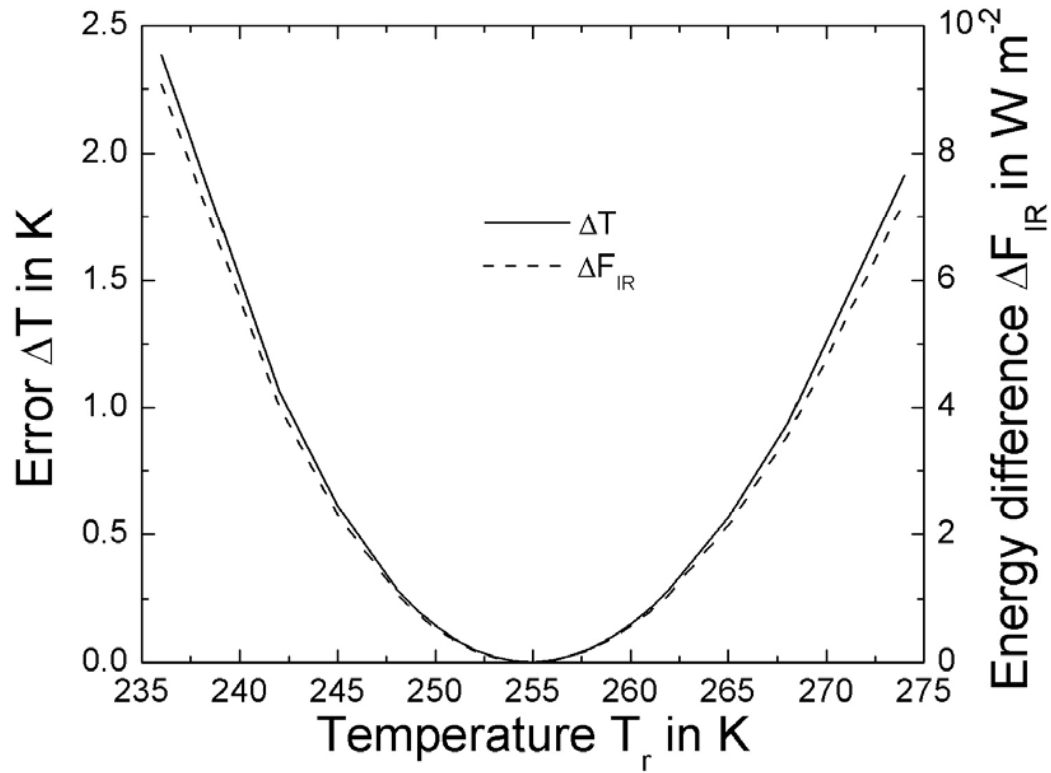

**Figure 8:** Error $\Delta T = T_s(\infty) - T_e$ versus reference temperature $T_r$ used by the linearization of the power law of Stefan [13] and Boltzmann [14] (see Eqs. (2.8) and (2.9)). Also shown is the corresponding energy difference $\Delta F_{IR}$. For the purpose of comparison: The total net anthropogenic radiative forcing illustrated in Figure 4 amounts 1.6 W m$^{-2}$.

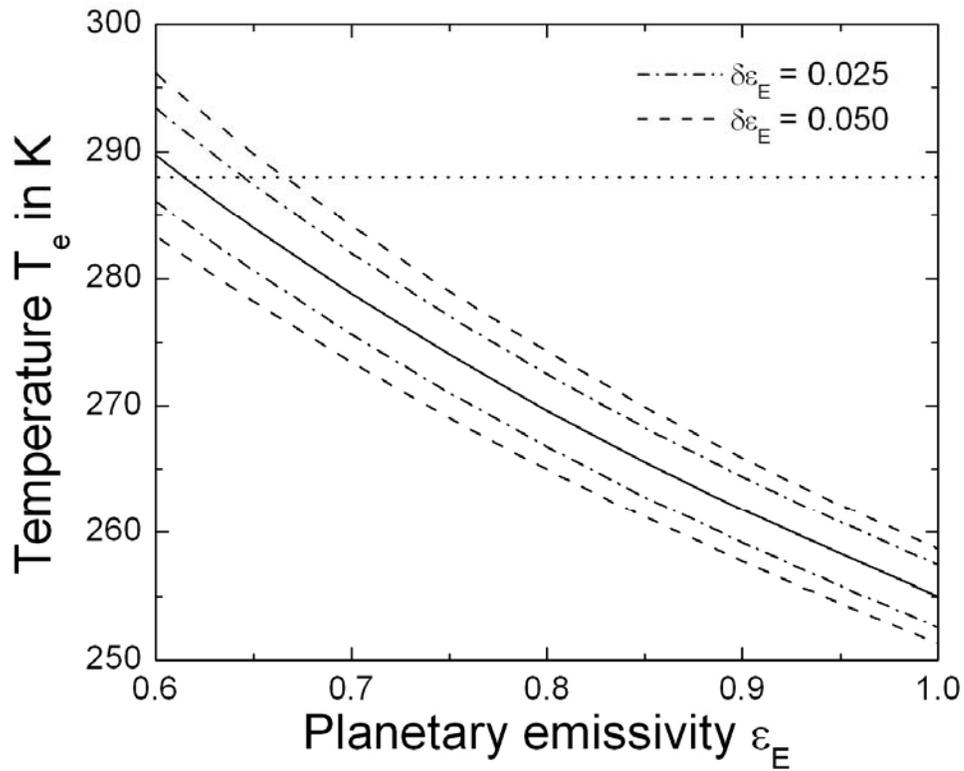

**Figure 9:** Temperature $T_e$ of the radiative equilibrium as a function of the planetary emissivity (solid line). Also shown is the mean global near-surface temperature (dotted line) and the range of uncertainty with which $T_e$ is fraught. The uncertainty was obtained using $\delta\alpha_E = 0.02$, $\delta S = 1.5 \text{ W m}^{-2}$, $\delta\varepsilon_E = 0.025$ and, alternatively, $\delta\varepsilon_E = 0.05$.

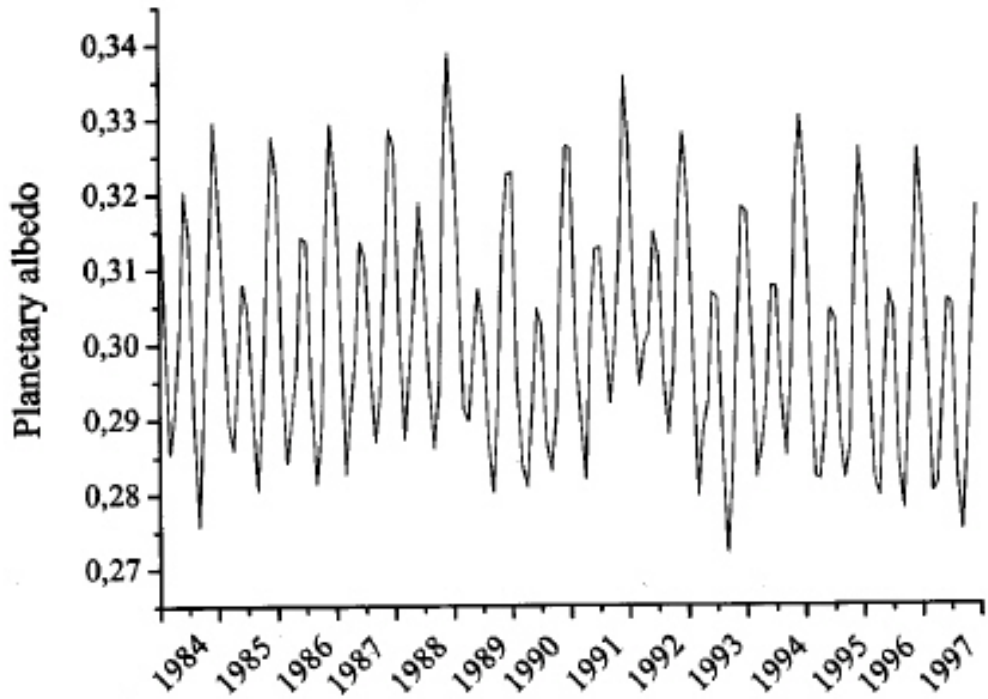

**Figure 10**: Long-term (1984-1997) time series of monthly averaged planetary albedo (adopted from Vardavas and Taylor [33]).

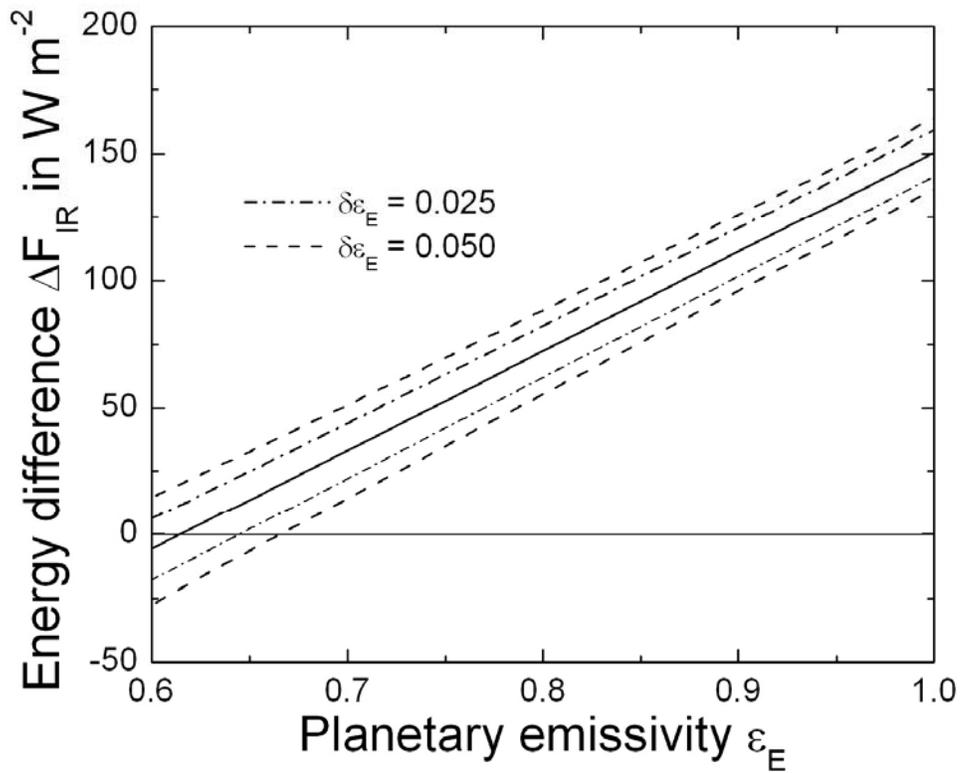

**Figure 11:** Energy difference $\Delta F_{IR}$ as a function of the planetary emissivity $\varepsilon_e$ (solid line). Also shown is the range of uncertainty with which $\Delta F_{IR}$ is fraught. For the purpose of comparison: The total net anthropogenic radiative forcing as illustrated in Figure 4 amounts 1.6 W m$^{-2}$.

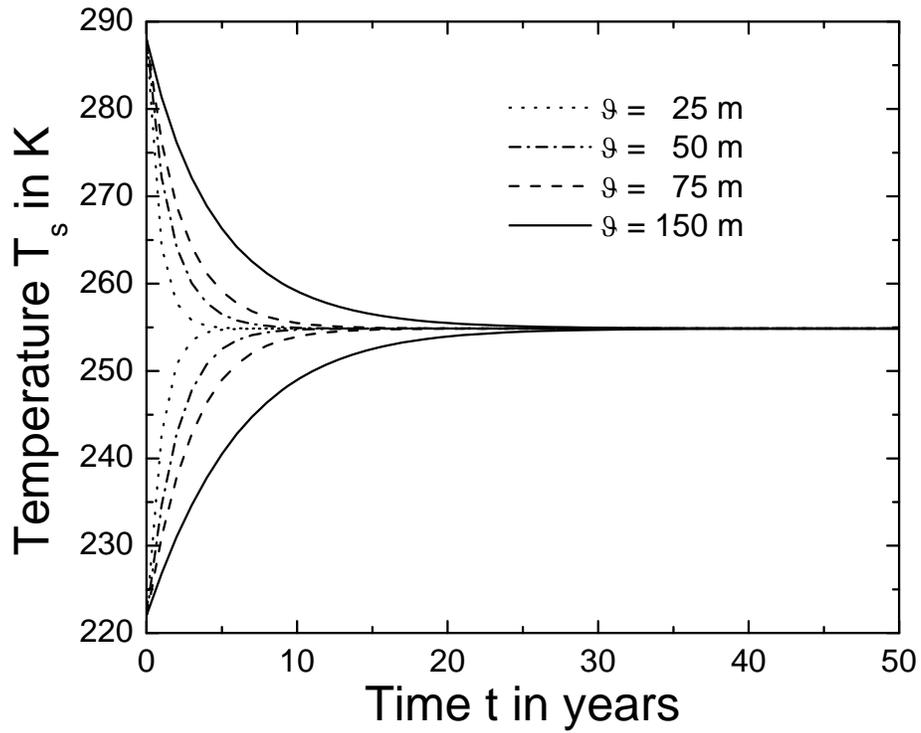

**Figure 12:** Convergence behavior of the numerical solution of Eq. (4.1) for four different values of the water layer thickness $\vartheta$. The initial conditions were assumed as $T_{s0} = 222 \text{ K}$ and $T_{s0} = 288 \text{ K}$.

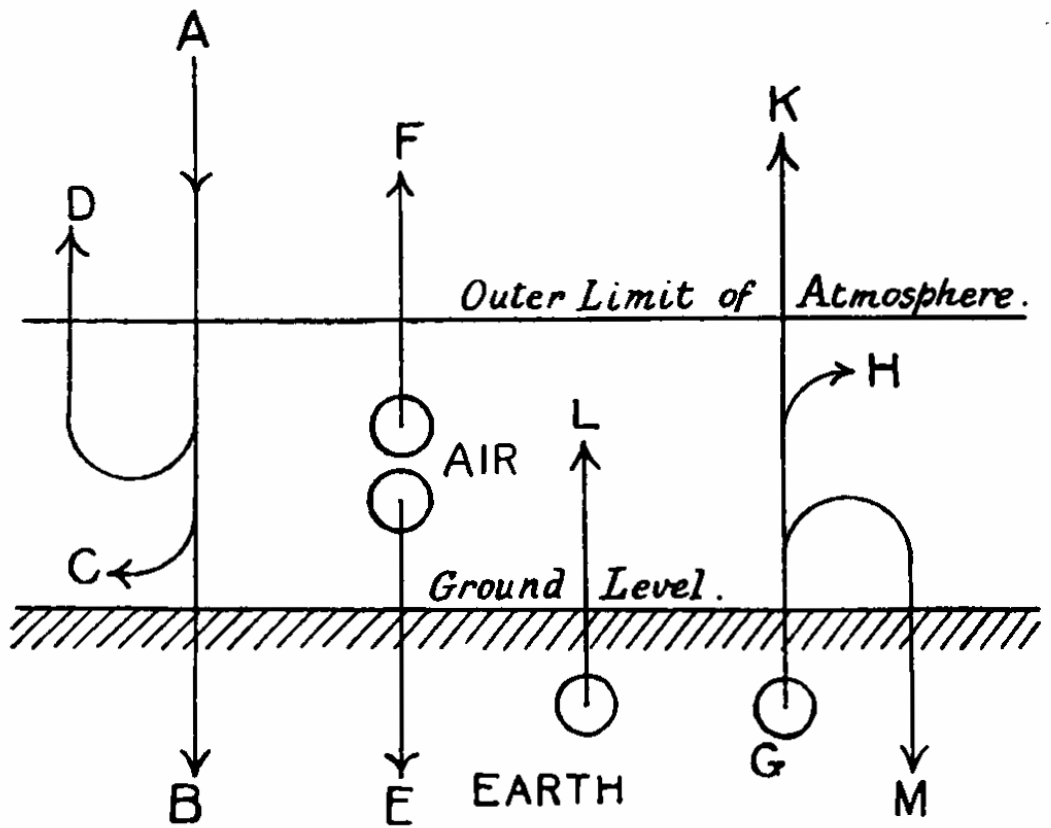

**Figure 13:** Dines' [58] sketch of the radiation balance of the atmosphere. The author described his sketch as follows: Radiant energy, denoted by A, reaches the atmosphere; of this a part (D) is reflected unchanged by the earth or air, a part (C) is absorbed by the air, and a part (B) is absorbed by the Earth. Meanwhile the Earth is radiating its heat (G) outwards; of this let M be reflected back, let H be absorbed and K transmitted. The air is also radiating downwards and upwards; let us call the amounts E and F, and if any part of E is reflected away by the earth it may be included in F. Also heat may pass from the earth to the air or in the opposite direction otherwise than by radiation; let us call this L, earth to air being the positive direction. According to Möller [59] the quantity M is based on Dines' mistake.

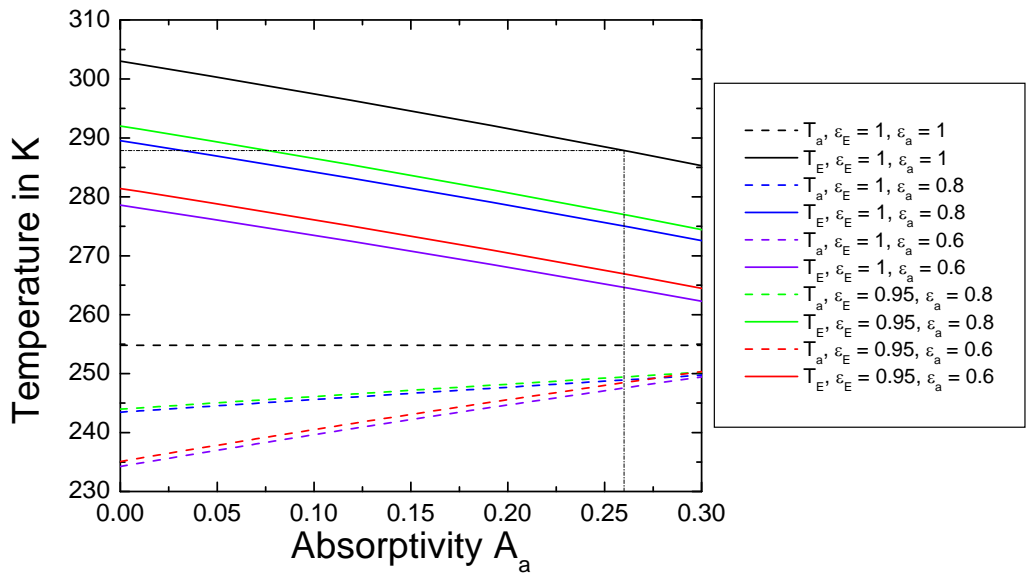

**Figure 14:** Uniform temperatures for the Earth's surface and the atmosphere provided by the two-layer model of radiative equilibrium versus absorptivity $A_a$ (adopted from Kramm et al. [60]).

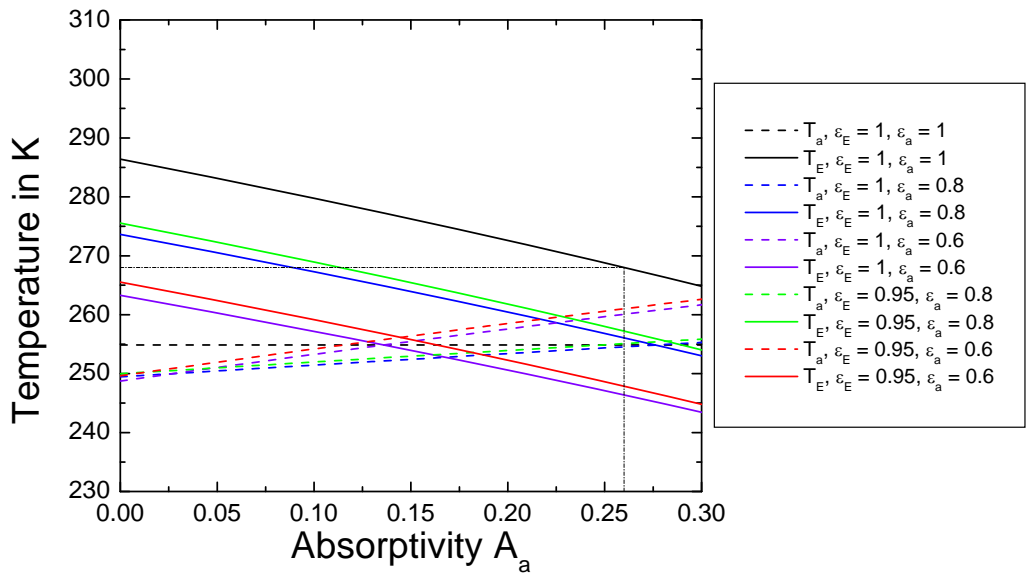

**Figure 15:** As in Figure 14, but the radiation flux balance at the Earth's surface is replaced by an energy flux balance including the fluxes of sensible and latent heat as suggested by Trenberth et al. [27] (adopted from Kramm et al. [60]).

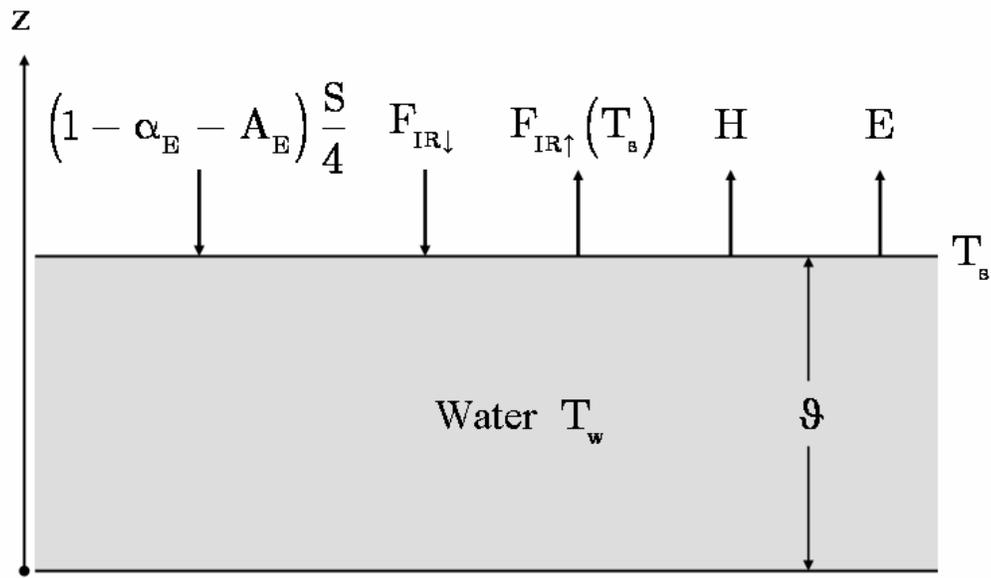

**Figure 16:** Sketch for deriving Eq. (2.1). The symbols are explained in appendix A.